# EVIDENCE OF MINOAN ASTRONOMY AND CALENDRICAL PRACTICES

Marianna Ridderstad

Helsinki University Observatory, P. O. Box 14, FI-00014 University of Helsinki, Finland
(ridderst@kruuna.helsinki.fi)

**Abstract**

In Minoan art, symbols for celestial objects were depicted frequently and often in a religious context. The most common were various solar and stellar symbols. The palace of Knossos was amply decorated with these symbols.

The rituals performed in Knossos and other Minoan palaces included the alteration of light and darkness, as well as the use of reflection. The Minoan primary goddess was a solar goddess, the 'Minoan Demeter'.

A Late Minoan clay disk has been identified as a ritual calendrical object showing the most important celestial cycles, especially the lunar octaeteris. The disk, as well as the Minoan stone *kernoi*, were probably used in relation to the Minoan festival calendar.

The orientations of the central courts of the palaces of Knossos, Phaistos, Mallia and Gournia were to the rising sun, whereas the Eastern palaces Zakros and Petras were oriented to the southernmost and the northernmost risings of the moon, respectively. The E-W axes of the courts of Knossos and Phaistos were oriented to the sunrise five days before the vernal equinox. This orientation is related to the five epagomenal days in the end of a year, which was probably the time of a Minoan festival.

One of the orientations of the Knossian Throne Room is towards the heliacal rising of Spica in 2000-1000 BCE. Spica rose heliacally at the time of vintage in Minoan times. The time near the date of the heliacal rising of Spica was the time of an important festival related to ctchonic deities, the Minoan predecessor of the Eleusinian Mysteries.

The myths of Minos, Demeter and Persephone probably have an astronomical origin, related to Minoan observations of the periods of the moon, Venus and Spica. These celestial events were related to the idea of renewal, which was central in the Minoan religion.

# 1. Introduction

The Minoan culture on the island of Crete was highly developed, with a writing system, central administration and a strong position in the sea trade of the Mediterranean. However, not much is known about the astronomy practiced by the Minoans or about their calendrical system. The Minoans must have been aware of many of the scientific developments originating in Egyptian and Mesopotamian cultures, and it is likely that they had developed as advanced a calendrical system as their contemporaries, but it has been difficult to try to settle the question on the basis of the scarce evidence available.

In this paper, the astronomy and the calendrical system of the Minoans are discussed in the light of literary, archaeological and archaeoastronomical evidence. The existing literary sources on Minoan astronomical practices are examined, including the Greek and Mycenaean calendars, the myth of Minos, and the role of celestial bodies in the Minoan religion based on comparative mythology. The Minoan astronomical practices are considered in the light of archaeological evidence, including Minoan art. New evidence of the Minoan calendrical tradition is presented in the form of a possible Minoan calendrical disk. Studies of the astronomical orientations of Minoan buildings and graves are reviewed, and new results on the orientations of Minoan palaces are presented. Finally, a possible origin for the Eleusinian Mysteries and Thesmophoria, based on the new archaeoastronomical evidence presented, is suggested.

## 2. Traces of Minoan astronomy in ancient literary sources

### 2.1 Minoan and Mycenaean Crete

The question of the differences between the central characteristics of the Minoan and the Mycenaean cultures is complicated, since the time of the beginning of the Mycenaean rule on Crete is not known exactly, and the Minoan influence on the Mycenaean culture was substantial.

It is important to relate any existing ancient literary, archaeological and archaeoastronomical evidence on the Minoan culture to the archaeological finds from the period before the Mycenaean rule on Crete. Hovever, the exact moment of the start of the Mycenaean influence is still a debated issue. Usually, the start of the Mycenaean period is taken to be about 1450 BCE at the earliest (see, e.g, Castleden 1990:34).[1]

The orientations of shrines and graves on Crete seem to have changed from the Minoan to Mycenaean times (Blomberg and Henriksson 2005). Mycenaeans on the mainland did not generally orient their graves, whereas Minoans did orient their graves, as well as their other buildings, mainly to east. There is, however, evidence that the Mycenaeans, who came from the Argolid, oriented their shrines and graves towards sunset, not only on the mainland, but also on Crete (Blomberg and Henriksson 2005 and references therein).

---

[1] This date corresponds to the start of Late Minoan II (LM II) in Evans' chronology. In this paper, I mostly follow this chronology based on the dating of Arthur Evans and used in his six-volume work *The Palace of Minos*. However, when presenting results given by other researchers in their papers, I use the naming convention of those particular papers.

With increasing Mycenaean presence on Crete, all other palaces except Knossos lost their importance, as did most peak sanctuaries, and cult rooms were situated in settlements (Whittaker 1997:46). It is difficult to deduce whether this was the result of a Mycenaean invasion, or whether there had been a Minoan upraisal or a series of natural disasters. There was a change in the objects placed in the shrines: female figures with their arms raised became more common (Whittaker 1997:42-6). The most important deities seem, however, to have been female, as before.

The Mycenaean society was a patriarchal society ruled by a king, and the most important female seems to have been a high priestess (Castleden 2005:78, 169). In the Minoan society females had been more prominent than men, both as deities and as performers of rituals (Evans 1930:227). Therefore, imposing the Mycenaean rule on the Minoan religious administration may have erased some of the earlier importance of women relative to men. There is evidence that the Minoans had both priestesses and priests (Castleden 1990:175), but whether the most important ruler of the Minoans before the arrival of Mycenaeans was female or male cannot, at the moment, be ascertained.

Castleden (1990:34) suggested that the Mycenaeans might have taken over Crete without much opposition. Certainly, taking over the Cretan religious rule rather unchanged would have been the most beneficial way for them, and there is evidence of this type of development (Blomberg and Henriksson 1996; Whittaker 1997:60-2).

## 2.2 Origin of the Hellenic lunar calendar

Unfortunately, no written sources on the calendars of the Minoans have been preserved. The written records left by the Minoans are, with the exception of some short religious formulas, solely concerned with the palaces' bookkeeping tasks. As it is known that the papyrus plant was very important for the Minoans, there must have been some records written on papyrus, too, but these have not been preserved. At present, to address the question on Minoan calendrical practices, one must rely on indirect evidence. Some clues are given by the Greek calendrical system and the traces of the Mycenaean ritual calendar.

The Hellenic calendars, of which the Attic calendar is the best known, were lunisolar calendars. In the Athenian festival calendar, the year began with the observation of the first new moon after the summer solstice (Samuel 1972:57). In other city-states, the year could begin, e.g., in midwinter. Each year had 12 synodic months, which resulted in 354 days total, which is 11 days short of one tropical year. To prevent the months moving backwards, intercalation was used, and occasionally a 13th month was inserted.

It is not known when the Greeks started to use the octaeteris-based, 99-month lunar calendar, although this must have been before 432 BCE (Samuel 1972:39). In this system, a 13th month was inserted every two or three years, resulting in the error of 1.6 days between the lunar calendar and the tropical year every eight years.

It has been suggested that the 99-month calendar was of Minoan origin (Blomberg and Henriksson 1996). It is known from the month names appearing in Linear B tablets that the Mycenaeans used a lunar ritual calendar, where some of the names of the months were based on the names of divinities, as in the later Greek calendars (see Samuel 1972:64-5 and refs. therein). One of the month names of Knossos tablets was used in Arcadia in Classical times (Chadwick 1958:128). The existence of a Minoan lunar calendar based on the eight-year lunar cycle is supported by archaeoastronomical studies (Blomberg and Henriksson 1996; 2001).

## 2.3 Minos *Enneoros* and the Minoan religious governmental system

The Greek mythology contains motifs, which hint at astronomical observations carried out for calendrical and religious purposes. Since some aspects of Greek religion probably had its origins in Minoan religion (Dietrich 1974 and refs. therein; Nilsson 1950), the Greek myths, too, can be used to find traces of Minoan astronomy.

One of the most important pieces of evidence concerning the Minoan ritual calendrical system is the mythology related to king Minos. According to Homer, Minos was the king of Crete, and, after his death, became one of the gods of the Underworld (*Odyssey* 11:568-71). Homer calls him "Minos *enneoros*" (*Odyssey* 19:178-80), which can be translated as "Minos, who reigned for nine years" or "Minos, who reigned as he was nine years old". As the concept of a periodically reigning king is familiar from Sparta and Mesopotamia (Frazer 1922: Chapter 24.3), the first option seems more likely. Although Homer tells about the Mycenaean Crete, not Minoan, the Mycenaean rule on Crete may have been close to the original Minoan religious governmental system.

The assumption that the Minoan rule was essentially based on their religion comes from the fact that the 'palaces' resemble more temple complexes than fortresses, and no other structures of the same size and importance have been found (Driessen 2003 and refs. therein; MacGillivray 2003). It is not known whether the Minoans had a king or a queen, or whether they were ruled by a high priest and/or priestess. The sacral nature of the power of the rulers of the palaces is, however, evident (Whittaker 1997:36-7 and refs. therein; Hitchcock 2003). It seems probable that the Minoan 'royalty' consisted of priests and priestesses, who may have acted in the epiphanies of deities (Castleden 1990:141 and refs. therein). Perhaps the system was close to the Mesopotamian one, where the high priest held the sacral kingship and was accompanied by the high priestess, who was the earthly manifestation of the great goddess (see, e.g., Klein 1992 and refs. therein). The Mycenaean king then would have taken over the same tasks as the earlier Minoan periodical sacral king.

In the Minoan art, there are scarcely any males depicted in leading positions (Castleden 1990:140). However, Minos having been a periodical priest-king would explain why he was not personally important, unlike pharaohs in Egypt. Every ninth year (which corresponds to once in every eight years if inclusive counting is used),[2] Minos went into a sacred cave to meet Zeus and to bring back new laws (Homer, *Odyssey* 19:172-178; Plato, *Laws* 1:624-25). This has been interpreted in the context of the need of a renewal of the king's power after a certain period. Same kinds of periodical renewal ceremonies of the power of the ruler were used both in the contemporary Egypt (the sed festival) and Babylonia (Frazer 1922:251-53). The model for the Minoan-Mycenaean kingship could thus have been taken from the sacral kingship practiced in Egypt, Mesopotamia, Anatolia, or Syria (see Castleden 1990:141).

Most of the legends concerning the rule of Minos are related to death, laws and bulls, thus hinting at sacrificial practices being a central part of the tasks of the religious administration. In the well-known legend of Theseus, Athenian boys and girls were every nine years sent to Knossos to be fed to the bull-monster Minotaur (Plutarch, *Life of Theseus* 15:1). This legend is reminiscent of the bull-leaping games depicted in Minoan art, and may have some historical basis, as indicated by the finds by Wall et al. (1986).

Minotaur was called Asterios, i.e., 'ruler of the stars' (Apollodorus, *The Library* 3:8-11), like Asterios the king of Crete, who raised the children of Europa and Zeus the Bull as his own (Diodorus

---

[2] Inclusive counting was used in Greek and Roman world, which sometimes lead to confusion (see Worthen 1991). This is easy to understand, as there was no concept of zero: the first year would be number one, which would lead, e.g., to the next cycle beginning on year nine of an octaeteris.

Siculus, *The Library* 4:60:3). It is tempting to see the Minotaur myth, which tells about the relationship between Pasiphae (who was later seen as a moon goddess, see Section 2.4.1) and the sacrificial bull of Minos, as a remnant of the sacred marriage of the Minoan high priestess and the priest-king, who was also the earthly representation of the lunar or solar bull. Castleden (1990:129-30) saw the Minoan "Bull God" as a solar god, but there is also evidence that strongly suggest the lunar interpretation (see Section 2.4.2).

If Minos was originally a lunar deity or the title of its priest, the rather curious idea of a mortal king becoming a judge in the underworld could be explained by the much older Mesopotamian myth of the moon god Nanna (Sin), who once in a month acted as a judge in the underworld, and was substituted by his three siblings to be able to leave there (Kramer 1961:43-6; Kramer 1963:132, 135, 146-7, 210).

Based on the eight-year ruling period of Minos and their studies on the orientations of Minoan buildings, Blomberg and Henriksson (1996) suggested that the Minoans used a lunisolar calendar based on the 99 synodic periods of the moon, i.e., the lunar octaeteris. They also suggested that the Minoan year probably began in the autumn. There is evidence of a tradition of *hieros gamos* on Minoan Crete (Ramsay 1912; Nilsson 1950:403, 550-53), and it was suggested by Koehl (2001) that this took place in the new year's celebration in the autumnal equinox.

It may be that in Minoan-Mycenaean Crete, the king was replaced, or his power was renewed, every eight years, when a festival with bull-leaping games and sacrifices were held. This festival would have started a new period of time, and would then have had strong ritual connections to renewal and fertility. A central part of this Minoan religious festival may have been a *hieros gamos* between the high priestess and the bull-king representing celestial bodies, perhaps the sun and the moon. The ritual coming-together of the most prominent celestial bodies, repeated in the union of the rulers, would then have marked the renewal of the king's power for the next octaeteris.

There are astronomical motifs in the Minoan art that hint at the role of celestial bodies in the sacrificial ceremonies (see Section 3.1). The archaeological evidence from the era before the Mycenaean rule on Crete does not, however, give clear indications on the identity of the ruler. The above picture is essentially a Mycenaean one in that it emphasizes the role of Minos as a male regent associated with the sun or the moon. There are alternatives to this view, based on, e.g., what is known of the cultures of Asia Minor, presented below in Section 2.4.2.

**2.4 Role of celestial bodies in Minoan religion**

As presented above, it is believed that the power of the Minoan ruling class was likely based on religious authority, and that the palaces were as much temples as the dwelling places of the administrators.

In addition to the palaces, caves, peak and spring sanctuaries were used as places of worship. In the Neopalatial period, the peak sanctuaries seem to have been closely connected with the central religious administration (Whittaker 1997:37-8). Some domestic shrines have been discovered.

All that is known about the Minoan religion comes either from archaeological finds or much later Greek literary sources as indirect evidence. The role of the king as it probably was at least in the Mycenaean era has already been discussed above. In Minoan times, however, all archaeological evidence points towards females being central both as deities and in performing the rituals (Evans 1930:227; Castleden 1990:175). The rituals included sacrifices, libations, (votive) offerings, communal feasting, bull-jumping sports, dancing, and probably also singing, burning sacrifices, and the epiphanies of deities (Marinatos 1986; Castleden 1990:53-62, 123-156, and refs. therein; Driessen

2003). The greatest communal activities were centered in the palaces (see Section 4.2.1).

Not much is known about the Minoan pantheon. It included many different goddesses or aspects of a single goddess, and a few male deities. There were at least the "Great Mother", the "Mistress of Animals", the "Dove Goddess", the "Poppy Goddess", the "Snake Goddess", a young male "Year-Spirit", the sacred bull, and animal-headed *genii* (Castleden 1990:124-25, 127-30, 142-3).

Minoan deities had clearly recognizable symbols, which could be used to denote the deity, or divinity in general. These included a poppy flower, a lily, a dove, a snake, a double axe, an "eight-shield", a pillar, a tree, a sacred garment, a sacral knot, and a "star" (see Evans 1921:430-40, 447, 1930:314-17; Nilsson 1950:155-340; Marinatos 1986:51-72).

The 'Dove Goddess' and the 'Snake Goddess' could be identified with almost all of the great goddesses of Classical Greece. Nilsson (1950: 488-91) argued that the goddesses, who were the protectors of the Greek cities in the Classical era, were associated with birds and snakes - an association which may be of Minoan origin. The birds and the snakes were related to the celestial and chthonic aspects, respectively, of Minoan goddesses (Evans 1902:85-7, 1921:508; Dawkins 1903:223; Castleden 1990:129).

The poppy flower of the Minoan 'Poppy Goddess' was associated in Classical Greek art with many goddesses, but, especially, it was the symbol of Demeter, who as the great mother and fertility goddess had a cult that had its origin in Minoan-Mycenaean times (Nilsson 1940:46; 1950: 403, 520-3; Coldstream 1984; Owens 1996a). The poppies hint at the use of narcotic substances in the rituals (Askitopoulou et al. 2002). Thus, the original 'Poppy Goddess' of the Minoans, as well as the Greeks, was Demeter (Theocritus, *Idyll* 7:157), i.e., the great mother goddess, whereas birds and snakes should be seen more as the emblems of the celestial and chthonic aspects of all Minoan goddesses.

It is not known exactly which Minoan divinities were associated with celestial phenomena, but these must have been some of the important ones, since symbols for celestial bodies were often pictured in the art, like in the art of Mesopotamia (see Section 3).

The complexity of the Greek mythology as it was presented already in Homeric times was due to many local gods and goddesses, many of which had had their origin in the Minoan religion, merging with the Mycenaean pantheon and with the later Greek influences. Therefore, some of the later Greek deities can give hints on which of the Minoan gods and goddesses were related to astronomical phenomena.

*2.4.1 Moon and Venus*

The crescent moon is not very often depicted in Minoan art (see Section 3.1), which makes the identification of the Minoan lunar deity difficult. A female lunar goddess is suggested by some representations of Minoan art (Blomberg and Henriksson 1996). Evans (1901) showed that the lunar deity was closely associated with the cult of trees, pillars, the horns of consegration and the double axe symbol.

The original Greek lunar goddess was Selene or Artemis. This confusion could arise from the fact that one of the Minoan principal goddesses was associated with the moon. The Classical and Mycenaean Artemis was probably the same as the Minoan 'Mistress of Wild Animals' (Castleden 1990:127). The Classical Greek literature mentions two Cretan goddesses, Britomartis and Diktynna, who had many similarities with the Greek Artemis (Nilsson 1950:510-3).

The important role of the moon for the Minoans is hinted at in the myth of Pasiphae, the wife of Minos. The name Pasiphae means "the all-shining", which is an epithet for the moon goddess, which she was (Pausanias, *Description of Greece* 3:26:1). In one version of the Minotaur myth, Pasiphae was supposed to make offerings to Aphrodite (Hyginus, *Fabulae* 40), which may reflect her connection to Venus, as both the moon and Venus have an eight-year cycle.

It is likely that the Minoans had one of their principal goddesses associated with Venus. In Linear A inscriptions, there appears the word *a-sa-sa-ra*, which is generally assumed to be the name of a goddess (see Younger 2000 and refs. therein). "Asasara" is probably related to the northwestern Semitic Astarte, who had her origins in the Mesopotamian goddess Inanna-Ishtar (Owens 1996b).

Already the Sumerian great goddess Inanna had been mainly associated with planet Venus. Her symbol was the eight-pointed rosette or star, and she visited the underworld, a story that can be seen as an allegory for the eight-day period of invisibility of the planet Venus, when it is in front of the sun before appearing as a morning star. The first recorded observations of Venus are from 1700-1600 BCE (the Ammisaduga tablets; see, e.g., Reiner and Pingree 1975). However, the first occurrences of the symbol of Inanna, an eight-pointed rosette, are Sumerian and much older, preceding 3000 BCE (van der Mierop 2007:50). In Babylonia and Assyria, Inanna was called Ishtar. Both Inanna and Ishtar were pictured and described as having horns, which may imply the crescent shape of Venus (Pannekoek 1961:35). Later, as the Semitic Astarte, Inanna-Ishtar retained her connection with planet Venus. The Hurrians/Hittites worshiped her as Shaushga (Akurgal 1962:80). Many of her properties as the great goddess were inherited by and mixed with later great goddesses, like Kybele, Demeter, Isis, Artemis and, especially, the Cyprian Aphrodite. Aphrodite, who was Venus herself, and Demeter, whose festival Thesmophoria was linked to the appearance of Venus as an evening star (see Section 4.4), had the closest connection to Venus in Classical Greece. In Crete, Aphrodite was worshiped as Aphrodite Ariadne in Amathus, which Evans (1902:87) believed to be a remnant of an earlier cult of the Knossian "Dove Goddess".

As in Mesopotamia, the eight-pointed rosette so frequently encountered in Minoan art could be the symbol for planet Venus (see Section 3.1), and, thus, also the symbol of the goddess Asasara. When the Minoan religion was combined with the Mycenaean system of gods and goddesses, and subsequently transmitted to the later Greek culture, many of the properties of the original deities prevailed, although their original significance was largely forgotten and their properties were inherited by new divinities. Especially, the rosette symbol was frequent in Mycenaean and later Greek art, apparently used as a symbol for deities or divine beings.[3]

The Mesopotamian Venus goddess was always closely related to the lunar and solar deities (the Utu/Shamash-Nanna/Sin-Inanna/Ishtar triad). It is possible that there was a similar situation in the Minoan pantheon, as there is evidence, based on the Greek mythology and the symbolism in Minoan art, that the solar and lunar deities were closely related in the rituals (see Sections 2.3, 2.4.2, 3.1, 4.4). In this context, one should note the Minoan tripartite shrines, which may be connected to these kinds of triads of divinities.

*2.4.2 Role of the sun*
The Greek mythology gives hardly any clues to the position of the solar deity in the Minoan pantheon. For the Greeks, the most important divinity was a weather god. The role of the sun god in the Classical Greek religion seems not to have been much more important than that of the moon. There is evidence that the Mycenaean culture at its peak adopted a great deal of the Minoan religion, which means that it was probably different from the later Dorian religion. However, by the Classical times, the primary deity in Greece was neither solar nor lunar.

The original sun god of the Greeks was Helios, and in the earliest times the lunar goddess was his sister (Hesiod, *Theogony* 371-4). A later Greek solar god was Artemis' brother Apollo, whose temple was in Delphi. Apollo had obtained Delphi from Themis, or Gaia and Poseidon, who had possessed the oracle in earlier times (Apollodorus, *The Library* 1.4.1:58-9; Pausanias, *Description of Greece* 10:5:6). Evans (1928:833-34) pointed out that there was evidence of early cult of the double axe in Delphi. He argued that the original cult of the place had been that of a Minoan goddess and her consort; in later times, they had been replaced by Apollo and Artemis (Evans 1928:833-44).

---
[3] Examples can be found in any source of Greek art; especially, the treatments of Greek pottery from different periods can be used. See, e.g., Nilsson 1950:417-8; Cook and Dupont 1998, Chapters 6 and 8; Boardman 1974, 1975, 1998.

Apollo was one of the deities associated with the eight-year cycle. Every eight years in Thebes, a festival of Apollo Ismenius was held, where representations of the sun, moon, and stars were carried in procession (Olcott 1941:242).

Dionysos, who was originally a dying and resurrecting Minoan vegetation god, had preceded the presence of Apollo in Delphi (Nilsson 1950:564-76). In the form of Iakkhos, the young Cretan Dionysos was connected to the myth of Demeter and Persephone (see the next Section). For the Greeks, Dionysos was the son of the moon goddess Semele (Homer, *Iliad* 14:323). Semele had 50 daughters, the Menae, who presided over the 50 lunar months, half of one octaeteris and one Olympiad (Pausanias, *Description of Greece* 5:1:4). Thus, in the Classical mythology it was the moon that governed the long calendrical cycles, and the annual vegetation cycle was in this way submissive to the lunar goddess, not to the solar god.

However, the Minoan religion probably differed in its central parts from the Classical religion. There has been accumulating evidence that the sun had a central role in the Minoan religion, a theory that is supported both by the Minoan archaeological finds and the orientations of Minoan buildings (Goodison 2001; Blomberg and Henriksson 2005 and refs. therein; see also Section 4).

The archaeological evidence of Minoan religion and some architectural features of the Minoan palaces have interesting parallels in the Hittite religion and temples. The Hittite temples resembled the Minoan palaces in their structure and probably also in function: great *pithoi* were stored in both (Evans 1928:269-70; Akurgal 1962:175). The most important deities of the Hittites were Teshup, a male weather god, and a female solar deity (Akurgal 1962:75-81). The Hittites worshiped not one, but two solar deities: goddess Arinna and the Sun of the Heaven. However, the Sun of the Heaven seems to have been a late import. Arinna could be related to Ariadne, the lady in the Knossian labyrinth, and, in other myths, the wife of Dionysos (Hyginus, *Fabulae* 42; Homer, *Odyssey* 11.320-5; Hesiod, *Theogony* 947-9). The original "Lady of the Labyrinth" was the great goddess worshiped at Knossos, *a-ta-na po-ti-ni-ja* of the Knossian Mycenaeans, as *potnia* means "mistress" (Chadwick 1958:125). Hicks (2002) argued that *a-ta-na* is related to Luwian *astanus*, the sun, as well as to the later Greek Athena.

The consort of Hittite Arinna had a double axe as his symbol. It is not known for certain, whether in Crete the double axe was the symbol of a female or a male divinity. Evans (1901:106-11) believed, based on parallels in Carian and Hittite cultures, that it was the symbol of a male deity. But the Minoan and the Mycenaean practices may have been different.

Since the double axe is depicted with the "horns of consegration", it has been seen as the symbol of Poseidon, the "Earth-Shaker" (Evans 1901:107; Castleden 1990:130, 135-6). The clear association of the double axe with the shrines of the Minoan tree and pillar cult point to the ancient Anatolian and Mesopotamian myth of the great goddess and her male companion in the form of a tree or a bull (cf. Kybele-Attis, Ishtar-Tammuz, also Aphrodite-Adonis; Evans 1901; Nilsson 1950:400-4 and Figs. 56, 61, 71-3). The youthful god, depicted with goddesses on the tree-shrine scenes of Minoan seals, had a central position in the Minoan religion as a male fertility god, a "Year-Spirit" (Castleden 1995:125-26). The sacred tree, the sacred bull, the young male "Year-Spirit", the Cretan Dionysos, Poseidon/Poteidan, and the deity of double axes, who dwelled in caves, may thus all have been aspects of the same Minoan god. This god, the 'Minoan Adonis' could have been the male partner of a Minoan female solar deity.

In Classical Anatolia, there was a cult of Mên Askaënos, who was a lunar god, and Demeter (Hardie 1912; Ramsay 1912). It was a mystery cult similar to the Eleusinian Mysteries. Although the goddess was called Demeter, she had much in common with Cybele, most importantly the fact that her partner was subordinate to her (Ramsay 1912). Mên was even referred to as Attis (Ramsay 1912:54). The name Mên obviously has a strong resemblance to Minos.

The tripartite shrines of Minoan Crete point towards three main divinities, the primary of which was likely female. But there are no clear indications, whether all of them were females. There certainly was an important male deity, but, on the other hand, many seals and frescoes depict only females, and more than two of them (see Section 3.1). Were these goddesses associated with celestial phenomena also? In Mesopotamia, the close partners of the main solar deity were the moon and Venus deities; this may have been the case with the Minoan deities, too.

The myth of king Minos points towards a priest-king being a representation of a celestial bull and the principal female deity also being a celestial deity. Evans (1901:168; 1930:457-58) seems to have supported the interpretation that the male bull-god was solar, although he believed that the goddess was the primary deity. Also Castleden (1990:129) suggested that the Minoan-Mycenaean Poseidon may have been a solar god. However, the Mycenaeans had a patriarchal society ruled by a *wanax*, whereas in the Minoan times, the role of women may have been more prominent: a sun goddess with a lunar bull as her male consort, resembling the Hittite pantheon. The myth of Iasion, the sacrificed consort of Demeter, indicates the submissive position of the male consort of a great Cretan fertility goddess (Homer, *Odyssey* 5:125-128; Apollodorus, *The Library* 3:12:1). Koehl (2001) pointed out that all representations of male-female unions on Minoan seals depict the female as the dominant partner.

For the Latvians, who preserved very ancient patterns of Indo-European mythology, the sun was female and her male consort was the moon, Menulis (Straizys and Klimka 1997). The Minoan pantheon was probably closer to the archaic Indo-European female-dominated pantheon than to the Greek male-dominated one, and one or more of the most important Minoan deities were female, accompanied by a fertility god.

At this point, it is impossible to reconstruct the Minoan pantheon with certainty, or to connect the celestial bodies with specific divinities. It can, however, be argued that the most important deity was likely female, and that the principal goddesses and gods included those associated with the sun, the moon, and fertility.

*2.4.3 Stars and asterisms*
The Minoan map of the sky is not known. The origin of Greek constellations was probably with Babylonian asterisms (Schaefer 2005). However, Roy (1994) suggested that the Babylonian constellations presented by Eudoxus, which were recorded a thousand years before his time, were originally observed on Crete.

By the Classical period, the Greeks had combined their pantheon and mythology with celestial phenomena to such an extent that not only did the sun, the moon, the planets, and the brightest stars have a divine meaning, but also every constellation on the sky was associated with a myth or a legend, often several. It is nearly impossible to extract any certain Minoan influences from this vast, multi-layered mythology. Some of the asterisms, which are associated with the relevant goddess mythology, are, however, worth considering. The most important of them is Virgo, which is associated with the myth of Demeter and Persephone.

*Virgo and Spica*
The constellation of Virgo has been associated with a maiden from ancient times (Allen 1963:460). Spica ($\alpha$ Virginis, 0.98 mag), Virgo's brightest star, is the "ear of wheat", which Virgo holds in her hand (Allen 1963:466). The connection of Virgo with the vegetation cycle mythology tells that the asterism may originally have had calendrical significance.

For the Greeks, Virgo was the asterism of Demeter's daughter, the virgin Persephone (Allen 1963:460-61). Many of the important goddesses associated with planet Venus (Ishtar, Kybele, Demeter, Isis), became at some point associated with constellation Virgo (Allen 1963:461-63), which probably resulted from the original connection of Virgo with a great goddess.

Demeter and Persephone have been suggested to be Minoan in origin (Castleden 1990:127 and refs. therein; Owens 1996a). Like Minos, they are closely connected with the number eight or nine. Demeter was also Thesmophoros, "the Law-Giver", which is another connection to king Minos (Callimachus, *Hymn 6 to Demeter*).

According to the legend, Persephone was to spend eight months of every year with her mother Demeter, and the rest of it, four months, in the underworld with Hades (Apollodorus, *The Library* 1:29-33). During the time when Persephone was in the underworld, the land went barren, and when she reappeared, grain could grow again, which means that she was essentially an agricultural fertility goddess.

The four-month period, when Persephone must dwell in the underworld, can be associated with the hot and dry Mediterranean summer months. However, in some versions of the myth, the treaty between Demeter and Hades was that Persephone were to spend only six months with Demeter (Ovid, *Metamorphoses* 5:564). This could correspond to the period from the heliacal rising of Spica in the autumn to its apparent cosmical setting in the spring. In Greece, Spica started to rise heliacally in the beginning of September around 2000 BCE, and by 600 BCE, it rose at the autumn equinox.

According to Nilsson (1940:55) the ear of corn displayed at the Eleusinian Mysteries can be equated with the "Corn Maiden", i.e., Persephone. Thus, the link between the date of the festival of an agricultural goddess and the constellation Virgo may have been important already to the Minoans.

However, September is too late for the grain harvest, and corresponds to the time of grape harvest instead. Hesiod (*Works and days* 609-17) tells us that around 700 BCE, the heliacal rising of Arcturus in the beginning of September marked the start of grape harvest. But in Minoan times, Arcturus rose earlier, and it was Spica that rose in the beginning of September. Thus, the heliacal rise of Spica may have marked the grape harvest for the Minoans, and Arcturus started to mark it only when Spica started to rise late due to the precession of the equinoxes.

The time of grape harvest preceded the time of sowing. The scenes on the Mycenaean and Minoan rings, where the Cretan Dionysos is depicted with a lamenting goddess, can thus be explained with this temporal connection (see, e.g., Nilsson 1950, Figs. 124 and 155). The heliacal rising of Spica marked the beginning of the grape harvest, and after that, when Virgo was fully visible, the autumn equinox, and the return of the Corn Maiden.

When Demeter looked for her daughter, she wandered the earth for nine days, and also the Eleusinian mysteries, held in honor of the two goddesses, lasted for nine days. According to the Attic calendar, the mysteries were celebrated in Boedromion in late summer, corresponding to about the 15th of September. This date corresponds to the heliacal rising of Spica near Athens in 1100 BCE. However, because of the lunisolar festival calendar of the Greeks, the date of the Eleusinian mysteries varied relative to our fixed solar calendar, and it can only be said for certain that it was celebrated in our August-September. Therefore, the heliacal rise of Spica could have coincided with the festival already in the Early Minoan (EM) period. Thus, the connection of Virgo and Spica with Persephone and Demeter may be related to celestial events originally observed by the Minoans.

*'The Torch-Bearer star'*
In the Greek mythology, Iacchus, who is equated with Dionysos (Strabo, *Geography* 10:3:10), is the torch-bearing young male companion of Demeter and Persephone in the Eleusinian mysteries

(Pausanias, *Description of Greece* 1.2:4). It is likened to a star bringing light to the mysteries (Aristophanes, *Frogs* 343). This could refer to Sirius ($\alpha$ Canis Majoris, -1.46 mag), the brightest star on the sky. Since in the earliest times, the planets were considered as wandering stars, the "star" could also have been Venus, which is the third brightest object on the sky after the sun and the moon. However, Venus was most often considered female in antiquity. Perhaps Iacchus was some other bright star, which had an astronomical connection to Venus or Virgo (i.e., Demeter and Persephone). One possibility is the star Arcturus ($\alpha$ Boötis, -0.07 mag), which was considered male in the mythology, and towards which some Minoan peak sanctuaries were oriented (Henriksson and Blomberg 1996; Blomberg and Henriksson 2001). From Minoan to Archaic times, the heliacal rise of Arcturus happened about ten days before the heliacal rise of Spica in the autumn, before the autumn equinox, which makes an interesting connection to the tradition that Iacchus as a torch-bearer lead the procession of the Eleusinian mysteries.

## 3. Archaeological evidence of Minoan astronomy

### 3.1 Astronomical motifs in Minoan art

Celestial bodies, or objects and symbols that can be interpreted as those, are a frequent motif in Minoan engraved and clay seals, moulds, rings, jewellery, garments, pottery, wallboards, ceiling boards, decorative inlays, frescoes, and reliefs.[4] Objects that seem to be moon sickles, rayed "suns", or stars, and even something that has been interpreted as the Milky Way by some scholars, are depicted (Evans 1902:88-94; Nilsson 1950:412-24; Goodison 2001 and refs. therein). Many of these are symbols, which in other contemporary cultures were commonly used for celestial bodies. In Figures 1-14, some of the Minoan celestial symbols are shown.

The most certain cases are those where the celestial bodies are pictured in apparently religious scenes with divinities or priestesses and priests. These scenes bear great similarity to the corresponding Mesopotamian and Cappadocian ones, where not only deities but also kings were pictured with the symbols for the sun, the moon, and Venus above them.[5] Early Babylonian seals have been found on Crete, one in a tomb from Early Minoan I, which proves that the Minoans were aware of this type of representations of deities from very early on (see Evans 1928:265, Fig. 158).

An object resembling a crescent moon is easily identified, and the identification is fairly certain. It is, however, rarer than the sun-like symbols, and two of the best examples are from Mycenaean cities (for these two gold rings, see Fig. 1 and, e.g., Nilsson 1950, Fig. 55). On a Cretan seal, a crescent moon is depicted on a shrine with horns and a tree (see Evans 1901, Fig. 59). The connection of the sickle-shape to the "horns of consegration", which are traditionally interpreted as bull's horns, was already noted by Evans (1901). It implies a connection between the Bull God and the Minoan lunar calendar. This connection explains the rarity of obvious lunar symbols: the very frequent depictions of the horns and the double axe also refer to the moon as the lunar deity.

---

[4] See, e.g., Evans 1902, Figs. 33, 42, 59, 62, 64; Evans 1925, Fig. 21, 26; Yule 1981, Plate 19: Motif 28: Stars; Olivier and Godart 1996, hereinafter CHIC, seals no. #004, #138, #222, #223, #247, #261, #304; Nilsson 1950, Figs. 38, 55, 71, 72, 83, 102, 130, 190-2; Castleden 1990, Figs. 1, 6-8; Fitton 2002, Figs. 40, 63, 99, 101, Plate 4. See also Figs. 1-10 and 12-14, which reproduce some symbols of Minoan art as presented in *Palace of Minos* I-IV by Evans 1921, 1928, 1930, and 1935.

[5] See, e.g., the Babylonian seal from 2400 BCE, where moon god is depicted (Mackenzie 1915, Fig. III1). The victory stele of Naram-Sin (2254-2218 BCE) depicts protecting symbols of the sun god above the scene (Mackenzie 1915, Fig. VI.2). Later, Babylonian *kudurru*s often depict the most important Mesopotamian triad of divinities: the sun god Utu, the moon god Sin, and the Venus goddess Inanna. Each of these divinities has the corresponding heavenly body as his/her symbol: a 'star' with four wavy rays radiating from it for the sun, a sickle for the moon, and an eight-pointed star for Venus; see, e.g., the *kudurru* of Ritti-Marduk for his services to Nebuchadnezzar I (1125-1104 BCE) in Mackenzie (1915, Fig. XIII.1). For the Cappadocian examples from 2400 BCE, see Evans (1930:205).

Another possibility is that, sometimes, the sickle could be planet Venus, since it has been estimated that it would have been possible for the ancients to observe its phases, especially the sickle shape of its brightest phase (van Gent 2004 and refs. therein).

The various rayed and circular 'solar' disks and star-shaped objects are found everywhere in the Minoan art. They are more difficult to categorize, not least because our modern perceptions of what are 'valid' solar and stellar symbols differ from those used in the antiquity. Perhaps the most famous example is the "sun", which could as well be a star, on the Minoan-style gold ring found in Mycenae (see Fig. 1). The scene on the ring shows the sun, the moon, and what looks like the Milky Way on the sky, as well as the "Poppy Goddess" seated under a tree, and two other female goddesses, one of which seems to have her young daughter with her. The ring may depict the lamenting of the Minoan-Mycenaean Demeter. The two female attendants of the goddess are represented with her in many other scenes, too (Evans 1928:339-42; Evans 1925, Figs. 13-15).

On a gold ring found in Kilia, a scene possibly related to the concept of Minoan *hieros gamos* is depicted: a man is standing in front of a house or a shrine, which has a tree growing on its top. He is reaching his hand towards a female, and a "sun" shines above them (Nilsson 1950:266, Fig. 130). Many Minoan seals depict a goat, a bull, or a *genius*, with a star (see Fig. 2). Star-shaped mason's marks have been found in the palaces of Knossos and Mallia (see: Evans 1921, Fig. 322; Nilsson 1950:242).

Very interesting celestial symbols are those found on a mould from Palaikastro. The scene pictured on the mould represents the "Poppy Goddess" standing with her arms raised, holding poppy seed heads (see, e.g., Nilsson 1950, Fig. 141). On the right side of her, there is a huge wheel-shaped solar disk or a star, and, on the left side of her, is a strange-looking round object with a stellar symbol surrounded by two circles of dots (Fig. 3). The object resembles the so-called Minoan cup-holed *kernoi*, in this case standing on a pedestal. There is a lunar crescent between the two dotted circles (Evans 1921:514), which points towards lunar or lunisolar interpretation of the use of the object. A similar object, painted on a jug from MM IB (Fig. 4), was, according to Evans (1935:94) likely connected to solar cult. The objects depicted on the mould and the jug, as well as real similar Minoan objects, may have had a ritual calendrical purpose, as will be shown in the next Section.

In addition to the clearly recognizable celestial bodies, there are symbols, which can be regarded as celestial, based on comparison with other cultures. The swastika, which is a solar symbol, is relatively rare in Minoan art. On one seal impression, of which seventeen examples were found, a goat is depicted with a swastika above it (Evans 1921, Fig. 372). A marble statue, which has the shape of a simple cross, a very old symbol for the sun or a star, was found among the cult objects of the Temple Repositories of Knossos (Evans 1902:90-94). The simple cross is also a frequent Minoan hieroglyph (CHIC:415-6). Another likely celestial symbol is the 'eye' with very prominent 'eye-lashes' (CHIC:387). The 'eye' is sometimes depicted as 'rising' over a (horizon?) line, which casts the doubt that it, too, is a solar symbol (CHIC #314). This may be related to the later belief of the sun as the eye of Zeus (Olcott 1914:288).

Evans (1921:514-17) interpreted the cross-signs found on Minoan objects as solar or stellar symbols; stellar when they were related to the cthonic aspect of the great goddess. He noticed that both in Knossos and in Mycenae, the griffin, which was the emblem of the goddess, was depicted with stars or asterisks (Evans 1921:549). The star sign also seemed to recur on the facades of shrines and on burial jars (Evans 1921:584).

The two most frequent symbols represented in Minoan art are the spiral and the "rosette", both of which are celestial symbols. They were used in Minoan and Minoan-Mycenaean art from the Early

Minoan Period to the end of the Late Minoan III (see Figs. 2 and 5-10). After that, they continued to be depicted in the Greek pottery, but decreasingly so towards the end of the Archaic period (see Boardman 1998).[6] Gradually, they got scarcer and seem to have been used more decoratively. In the well-known red-figure pottery style, developed in Athens around 500 BCE, only very decorative half-rosettes are seen frequently, and they seem to have lost their original connection to the celestial bodies (see Boardman 1974, 1975).

The spiral is an ancient solar symbol, which was frequent already in Egypt (Evans 1935:253) and in European megalithic art (see, e.g., Cope 2004:58-61). It is closely connected to the representation of the solar disk as concentric circles (Figs. 5 and 7, lower left). The spiral form can be taken to represent the apparent path of the sun from a solstice to solstice, if one thinks, as the ancients did, that the sun returned to its rising position during the night by taking a similar path in the netherworld (see Fig. 11). The two spirals opening in different directions thus correspond to the two routes of the sun in the sky: the increasing curve (from the winter solstice to the summer solstice), and the decreasing one (from the summer solstice to the winter solstice). These two spirals can be connected to form a double spiral (the 'S-form'), which is frequent in Minoan pottery and seals. Other spiral forms are the triple spiral or triskelion, and the quadruple spiral form (see Figs. 2, 5, 7 and 10).

That the spiral form was associated with the sun by the Minoans is indicated by its use on the façade of a tripartite shrine on the chlorite rhyton found in Zakros (see, e.g., Fitton 2002, Fig. 86). Bonfires were burned in many peak sanctuaries; e.g., at Traostalos (Chryssoulaki 2001), which was the peak sanctuary closely associated with the palace of Zakros (Fitton 2002:101). The association of bonfires with solar worship is an ancient one (Olcott 1914:231-48). Henriksson and Blomberg (1996) showed some of the walls of Traostalos were oriented towards the peak of Modi. However, the structure is badly preserved, and, based on the existing remains of the walls, and the sloping terrain towards east, the place could have been used to watch the summer solstice sunrise at (az 60 deg), too.

The recognition of the spiral as a solar symbol has been hampered by the fact that the spiral form in Minoan art can also be associated with the spiral form of some doliums in the "Marine style" pottery. The difference is, however, clear: the solar spirals clearly do not resemble doliums. One modern perception of spirals is to see spirals as waves, but, in Minoan art, they are never associated with the representations of sea water. Instead, the symbol with which the spiral is most often connected to is the rosette (see Figs. 7, 8 and 10).

The 'rosette', the flower-shaped symbol (e.g., Fig. 6, lower row, second from right), is very frequent in Minoan art, and is especially found in 'royal' objects (gold pins, necklaces) and in the palaces (frescoes, seals, *pithoi*, pottery). The rosette is a very old symbol, being already present in the megalithic art (see, e.g., Cope 2004:241). In many cultures, the rosette has been, and still is, a solar symbol, but the earliest certain meaning of it is found in Mesopotamia, where the eight-pointed rosette was the symbol of planet Venus (van Buren 1939, 1945). Thus, the rosette is a solar or stellar symbol, and closely connected to the cross-like or wheel-like celestial symbols.

The flower-like rosette usually has an even number of petals, often eight. The obvious celestial connection for the number eight is the octaeteris, the cycle of eight tropical years, which roughly corresponds to 99 lunations and five synodical periods of Venus. Another reason for the choice of eight is the eight-day period of invisibility of Venus between its periods of appearances as a morning and an evening star. This period appears in the myth of Inanna-Ishtar, a likely predecessor of Minoan Asasara. Much of the contemporary Mesopotamian astronomy was probably familiar to the Minoans. The symbols of celestial bodies may have been the same, too. A Middle Minoan III (MM III) mould found by Evans (1902:65; 1921:487-9) in the Temple Repositories of Knossos shows three symbols, which very much resemble the three Mesopotamian triad of the sun (the disk), the moon (the crescent), and Venus (the rosette).

---

[6] Examples of the symbols on pottery can be found on almost all figures in Boardman 1998, which is why I have not specified the figures.

The palace of Knossos was amply decorated, not only with nature scenes and double axes, but also with celestial symbols: rosettes, half-rosettes, spiraliforms, and *vesicae piscis* -like stellar representantions (Figs. 7-10; see also Section 4.2.1). Especially the large palatial *pithoi* and pottery were frequently decorated with rosettes from the EM to the LM period (Fig. 7, for more examples, see: Evans 1921: Figs. 80b, 186; Plates 1, 2; 1928: Figs. 205-06, 239, 244-45, 284; Plate 9; 1935: Figs. 177, 260-62, 268, 271, 273, 282-85, 297, 300-01), which would be obvious if the rosette were a religious celestial symbol related to the cult of the sun goddess. A frequent motif both in pottery and in wall paintings was the spiral, which had a rosette in the middle (Figs. 7-8). As the spiral was a solar symbol, the rosette could have been the stellar symbol of Venus, since it rises and sets close to the sun.

As the above examples show, the sun or a star was a central part of the Minoan cult, along with the "horns of consegration", which were frequent in shrines. A solar or stellar object is sometimes depicted in the middle of the double axe (see Fig. 7), in a way similar to the double axe being placed in the middle of the horns (Evans 1902: Fig. 70; 1921: Fig. 312c), indicating that their symbolism was related to the symbolism of the horns. Therefore, as the combined 'horns-and-axe' motif can be related to rituals where bulls were sacrificed, it is possible that the dates of the rituals were determined using a calendar based on the movements of the celestial body symbolized by the rosette, as well as by the moon, symbolized by the horns.

At this point, it cannot be determined for certain whether the rosette was the symbol of the sun or Venus. Perhaps the same symbols were used for the sun and a star (Evans 1921:514). However, to sum up, the above examples clearly show that celestial bodies were important in the Minoan religion.

## 3.2 Archaeological finds related to the Minoan calendrical system

### 3.2.1 Cup-Holed Stones (Kernoi) on Minoan Crete

To date, at least 167 different stones with cup-hole marks are known from Minoan Crete. They range in size from tens of centimeters to almost one meter, and the arrangement of the cup-holes varies. They have been found in public areas as well as in shrines, and have been interpreted as libation tables, *kernoi*, game boards, and some of them as calendars.

Evans (1930:390-96) found one *kernos* in a ritual area in the palace of Knossos, and interpreted it, as all similar cup-holed stones, as a pavement game. Also Hillbom (2005:82-3) interpreted most of them as game boards; however, he pointed out that some of them may have served other functions. The only certain example of a Minoan gameboard, the Royal Gameboard of Knossos, was connected to solar or stellar symbolism and the goddess cult already by Evans (1900:77-82, 1921:478-80, 485). The Royal Gameboard contains the symbolism of number eight.

Herberger (1983) interpreted the large so-called Mallia altar stone (Fig. 12), which has 33 small plus one large hole in a circle, as a calendar for the eight-year lunar cycle. He also discussed a stone with 13 cup-holes in a circle, found in Kato Chrysolakkos, and concluded that it shows the same cycle. However, as argued by Hillbom, he did not seem to be aware of the other circle on that stone, with 28 holes. Hillbom pointed out that the stone could have been a calendar of 13x28=364 days, but that it did not fit the 29.5-day synodic period of the moon (Hillbom 2005:82). However, he did not mention the other cycle of the moon, the sidereal period of 27.3 days. The sidereal month is observable in periods of three (3x27.32=82), when it closes at the same time of day, and could have been known to the

Minoans, as it was to the Babylonians from very early on. The Minoans used fractions in their bookkeeping, and so could have performed the calculation. Perhaps the stone was an attempt to represent the lunar sidereal year of 13x27.3=355, which is close to one lunar synodic year of 12x29.5=354 days.

In addition to the two stones discussed by Herberger and Hillbom, there is another large (40x50 cm) stone with a circle of eight holes known from Mallia (see Hillbom 2005:155). The eight holes could represent the octaeteris.

Of the 167 cup-holed stones catalogued in the doctoral thesis of Hillbom (2005:127-72), 92 are only partially preserved or, for some other reason, the original number of the cup-holes is uncertain. Most of those (26 stones), for which the exact number of the cup-holes in some formation (circle, rectangle, spiral, or other) is known, have either 12 or 13 cup-holes in a circle. These numbers have an obvious connection to the 12 and 13 months of lunisolar calendars.

Most of the 12/13-holed stones have been found in Phaistos or Vasiliki and dated to all periods from Early to Late Minoan. From Gournia, which is near Vasiliki, a stone with an inner half-circle with eight cup-holes and the outer full circle, with 33 holes, has been found (Hillbom 2005:140). These numbers can be related to the eight years and 99 months of one lunar octaeteris.

A clue to the purpose of the Minoan stone disks with cup-holes is the Palaikastro mould, discussed in the previous Section. It depicts a goddess with a circular object with two circles of holes on it. It seems likely that the object depicted is a religious object, a calendrical disk similar to the cup-holed stones and to the clay disk discussed in the next Section.

According to Hillbom (2005:83-5), the games, which supposedly were played using the cup-holes, were associated with the Minoan religion, like in Egypt, where the game *senet* increasingly acquired religious meanings over time. Therefore, whether the cup-holed stones were games or actual calendars, they could have been associated with important astronomical cycles and used for divination or for other mainly religious practices.

*3.2.2 Late Minoan Calendrical Clay Disk*

In his excavation of Knossos, Evans (1935:735-6) found a LM III clay disk, described as a "cheese-strainer" (Fig. 13). The disk is about 40 cm across and perforated: it has five concentric circles of small holes. The numbers of the holes from the center of the disk are 1, 15, 24, 38 and 61 (see Fig. 13). The function of the holes is unknown. A seven-rayed sun or a star is pictured in the center of the disk, indicating that it may be related to ritual calendrical purposes.

The holes in the disk, suitable to hold incense sticks or pegs, bring to mind the later Greek *parapegmata*, which were calendars with holes where a peg would be inserted to show the right day, week and/or month, along with star-phases and weather. The disk may be related to the Minoan cup-holed stone disks (see the previous Section).

If the disk were some kind of calendrical object, it could have been used as follows. The fifteen holes in the center of the disk could be half of the synodic month of 29.5 days, and the hole in the center of the disk could have been used to indicate, whether the month was supposed to have 29 or 30 full days, i.e., whether one hole was to be skipped.

The next circle has 24 holes. Together with the ring of 15 holes, these holes could have been used to count the length of one year of the contemporary Egyptian calendar, which was 15x24=360 days. The Minoans may well have used the same system as Egyptians with twelve 30-day months and a 360-day year, plus epagomenal five days in the end of the year. However, there are no markings indicative of periodical additions on the disk.

Using the disk one is also able to count 29.5/2*24=354 days, which correspond to 12 lunations,

11 days short of one tropical year. Similarly, the 38 holes of the third circle could be used to calculate 38 lunations. But a much better choice would have been 37 holes corresponding to 37 lunations, which are only 3.1 days shorter than three tropical years. However, together with the outermost ring with 61 holes, the 38-holed circle can be used to calculate one lunar octaeteris, which is 38+61=37+62=99 synodic months long.

In the later Hellenic calendar, the leap months were inserted at the ends of the 3rd, 5th, and 8th year of the lunar octaeteris, i.e., the 37th, 62nd, and 99th months were leap months. The second sub-period of 62 months differs from 5 solar years by 4.6 days only. If the first hole of the 38-circle is the starting hole, then, after 36 moves, a leap month would be inserted. A second peg, moved simultaneously with the first peg, but using the 61-ring, would be moved until the 62 months were full, the last of them being a leap month, and indicated by moving a peg to the inner circle of 38 holes. This peg would be then moved for another 37 months to continue to the full count of 99 months (the last of which is again a leap month). In this way, the holes of the disk could be used to track 99 lunar months, one octaeteris.

Based on the above consideration, a lunisolar calendrical interpretation for the disk seems likely. It is not a practical calendar with month names and numbers, but rather a ritual device, which concretely shows the basic calendrical cycles. It may have been used to hold incense sticks to be burned on important calendrical dates, or it may have been used for libations that were performed on astronomically important dates. The symbolism of the disk does not necessarily reveal the details of its use.

Although made of clay, the disk must be seen in the same context with the stone *kernoi*. The ritual calendrical interpretation is most probable for the large stone "offering tables", which were, like that of Mallia, obviously meant to be used permanently and in a religious context. It may be that various kinds of astronomically motivated 'calendar disks' were used in the palaces and shrines.

## 4. Minoan religion and calendar in the light of astronomical orientations

### 4.1 Orientations of Minoan buildings and graves

Most Minoan buildings with ritual functions seem to have been oriented towards east. The early studies on the orientations of Minoan palaces all concluded that the palaces were probably deliberately oriented towards the east and the rising sun (Marinatos 1934; Shaw 1977). Blomberg and Henriksson (2001 and refs. therein; 2005; 2006; 2008) showed the orientations of Minoan religious buildings (palaces, villas, and peak sanctuaries) to be towards major celestial events: the rising sun at equinoxes and solstices, one southernmost rising of the moon, and the heliacal rising and apparent cosmical setting of Arcturus. The palace of Knossos was also recently studied by Goodison (2001), who concluded that the "Throne Room" of Knossos was directed to the rising sun on all the main solar dates of the year. According to her, certain parts of the palaces may have been used as "theatres of the sun" in Minoan rituals.

Orientations of Minoan and Minoan-Mycenaean graves on Crete were measured by Papathanassiou et al. (1992), Papathanassiou and Hoskin (1996), and Goodison (2001 and refs. therein). These studies showed that in the Bronze Age Crete, the orientations of tombs were mainly towards the east.

### 4.2 Orientations of Minoan palaces

*4.2.1 Function of the palaces*

The Minoan palaces not only governed the surrounding region, collected taxes and lead the foreign trade, but were also places where the most important religious ceremonies of the Minoan society were carried out, and possibly dwelling places of the religious administration (Driessen 2003 and refs. therein; Hitchcock 2003 and refs. therein; MacGillivray 2003). Thus, they could be called "temple complexes" rather than palaces (Castleden 1989:70). The model for the palaces may have been taken from Anatolian and Mesopotamian temples, which were economic centres (Castleden 1990:6-7).

The four largest palaces were Knossos, Mallia, Phaistos, and Zakros. In addition, there were a number of smaller palaces: Gournia, Petras, and Galatas. There are probably more: Arkhanes, Palaikastro, Khania, Kommos, Monastiraki, Myrtos, Protoria, and Stavromenos have been suggested (MacGillivray 2003; Miller 2003). The largest of the palaces was Knossos, and it probably held a dominating position among the palaces from Late Minoan II on (Castleden 1990:23).

The most central place of a Minoan palace was the large central courtyard, where religious feasts, ritual dancing, bull-jumping, or other communal religious ceremonies probably took place (e.g., Driessen 2003). The central court was originally an open space, and only gradually was it encircled within the buildings of the palace complex (Driessen 2003).

In all four largest palaces, which are oriented N-S (Knossos, Phaistos, Mallia) or NE-SW (Zakros), the main cult rooms were located in the western wing, facing east, in the period 2000-1700 BCE, when the now visible phases of the palaces were constructed. Based on this, Shaw (1977) was the first to suggest the rough N-S orientation of the palaces to be connected to the function of the central court. He then argued, based on his measurements, that this orientation was due to solar and lunar orientations; a conclusion already made by S. Marinatos in 1934.

In Middle and Late Minoan Knossos, a tripartite shrine was facing the middle of the central court on the ground floor on the western side (Evans 1921:424:25). Next to it were the entrances to some of the cult rooms of the palace: the Temple Repositories and the Pillar Crypts, which actually were behind the façade of the tripartite shrine. The "Pillar Crypt" was a dark room with a pillar with double axe markings, where rituals of chthonic deities were likely performed (Evans 1901, 1928:322). The architecture of the pillar crypt was repeated in the room above it (Evans 1930:4). Marinatos (1993) argued that these two rooms were meant for rituals connected to darkness and light, respectively.

On the southern side of the Pillar crypts, there was the Corridor of the House Tablets, where Blomberg and Henriksson (2001) argued that calendrical observations had been carried out. The Corridor of the House Tablets had a stone basin, which may have been used to observe the sun indirectly, when filled with water (Blomberg and Henriksson 2001). Other stone basins had been found by Evans (1935:935-6) in the NW corridor leading to the central court, and in some nearby basement rooms.

In the NW end of the central court, there was the Throne Room complex (Evans 1899:35-42, 1935:901-921). An antechamber, which was separated from the central court by polythyrons, was, in turn, separated by polythyrons from the Throne Room. Castleden (2005:169) pointed out the plaster "throne" seat of the Throne Room was likely the seat of a priestess. The Throne Room had a so-called "Lustral Basin", which has been seen as an imitation of the natural caves, which were important cult places for the Minoans (Evans 1928:322). On the floor, there were six alabaster oil-containers, decorated with spirals and rosettes (Evans 1899:41).

The corridor going through the antechamber and the Throne Room ended to a wall in an inner sanctuary, which had a small shelf, possibly a place for a cult image (Evans 1935:910, 920).

In the western wing of the Knossos palace, there was a second "throne" seat in the Hall of the Double Axes. Also this room had an antechamber, which had eleven sets of polythyrons on its western, eastern and southern sides (Evans 1930:318-48). The room with a "throne" did not have a lustral basin, but, instead, ended in a light well, which had double axes carved on its walls, like those on the walls of the Corridor of the House Tablets.

The Hall of the Double Axe, as well as many of the nearby rooms, were decorated with the 'spiral and rosette' dado, indicating its connection to celestial deities (see Fig. 11, lower right). In this part of the palace, a Minoan *kernos* or game, with 12+3 holes, was found, carved on the pavement slabs (Evans 1930:390-1; see also Section 3.2.1).

Goodison (2001) argued that regulating the amount of natural light in the Throne Room was a central part of the rituals held within. According to her, the Throne Room complex was a kind of "theatre of the sun", where a person seated on the throne was illuminated by the rays of the rising midwinter sun through the southernmost doors of the anteroom. A similar effect was possible in the Hall of the Double Axes (Goodison 2001 and refs. therein). The equinoctial sunrise, in turn, would illuminate the Inner Sanctuary. At the summer solstice, the rising sun would illuminate the Lustral basin, and a person possibly situated in there. There is also a fourth alignment, which, according to Goodison, coincides with a large cluster of Mesara-type tholos tombs. Below I suggest an interpretation for this orientation.

The idea of the palace of Knossos as a "theatre of the sun" can be connected to the archaeological finds, which show that the ability of objects to reflect light was valued. On the floor of the Throne Room, Evans (1899:41; 1935:928-31) found cutted crystals, one of them crescent-shaped (Fig. 14), and remains of a crystal bowl. Inside the NE wall of the room, he discovered more crystals, as well as gemstones, gold foil, and "brilliantly colored faïence" (Evans 1935:934). He dated the deposit to MM IIIB. On the floor, he found faïence plaques, which were found upside down, and had linear markings for attaching them in correct positions (Evans 1899:41-2; 1935:40-1). They had probably been attached to the walls or to the ceiling of the room. They had been assembled to the form of *vesicae piscis*, leaving a star-shaped space in the middle (see Figs. 6 and 9, lower right). This pattern is also found in Minoan pottery already in EM II (Evans 1935:940), and on the famous Royal Gaming Board (Evans 1921, Plate V). The only other place in the palace where similar cutted and marked crystals and faïence plaques were found, was in the deposit in a corridor in the east wing, near the Hall of the Double Axes, where the other "throne" seat had been situated (Evans 1930:405-9). Moreover, Evans (1902:46-7; 1921:469-71) found, in the Temple Repositories, dated MM III, a crystal disk of 11 cm in diameter, backed with metal foil, that could have been used to mirror light. Crystal petals, which had probably been parts of rosettes of about 11 cm in diameter, and gold foil were found with the disk. They had probably been part of the inlay of a wall, a ceiling, a gameboard, or a lid.

All the evidence presented above strongly suggests that the "throne" room actually was the central shrine of the palace, where the high priestess appeared in the epiphany of the goddess at key solar dates of the year. The polythyrons would have been opened, and the sunlight let in to reflect from all the inlayed crystals, faïences, and metal foils, to create an effect of the goddess appearing in a revelation-like event in the middle of the bright light. The gold, gemstones, and crystals attached to the garment of the goddess would have enchanced the effect. Based on the finds, bowls of water and mirrors may have been used, too.

The two kinds of "throne" rooms and the Pillar Crypt indicate that the rituals performed in Knossos had a two-fold nature: one closer to the earth (the 'dark room'), as well as one closer to the sky (the 'light room'). Both spaces required the light of the sun to be either allowed to pour in or to be shut out, as is revealed by the polythyrons.

The palace of Phaistos, Mallia and Zakros have essentially the same features as Knossos, although the architectural solutions are different. Lustral basins, ceremonial spaces with polythyrons, and pillar crypts are all present. The polythyrons indicate that the alteration between light and darkness was used in the rituals performed in these palaces, too.

The palace of Mallia is the one resembling Knossos the most in its layout. It can be noted that in Mallia, the pillars in the Pillar crypt are not only engraved with double axes, but also with a star and a trident (Nilsson 1950:242). Mallia also has, on the SW side of the central court, the 34-cupholed Mallia *kernos* (see Section 3.2.1 above).

*4.2.2. Orientations of the palaces*
As the most important places for Minoan communal rituals were the palatial central courtyards, the orientations of them can give more reliable information on the original dates of the religious feasts than the surrounding constructions. In the following, the orientations of the central courtyards of six palaces (Knossos, Phaistos, Mallia, Gournia, Zakros, and Petras; see Table 1) are presented and discussed in the context of their possible astronomical significance. The remains of the palaces now visible date mostly from the beginning of MM III around 1700 BCE. The central courts of the palaces were, however, probably the same already in the beginning of Middle Minoan IB around 2000 BCE , when the large palaces were first constructed. The orientations were measured using a compass and the position of the sun, and the results were checked against results obtained using satellite photographs and site plans, as well as the results of previous studies, when these were available.

*Knossos*
The long axis of the central court of Knossos is oriented 10/190 deg N-S, thus deviating only 10 degrees from the meridian line. The present central court is from the Second Palace Period (built around 1700 BCE), but it was built above the earlier court, and the orientation stayed the same (Evans 1903:34-9; 1935:902-3).

In the north of the palace, the horizon levels off towards the coast. The Minoan town was located west of the palace. In the east, the Aelias ridge causes the horizon level to rise over 10 degrees. In the south, the prominent peak of Juktas is clearly visible. Scully (1979) suggested that the main axis of the palace was deliberately oriented towards it, and perhaps the peak sanctuary on the top of it. However, the central court is not oriented strictly to the mountain peak, but rather towards the east side of it. Therefore, it is to be suspected that there were other, more important directions, which principally determined the orientation of the court and the palace.

Marinatos (1934) suggested that the palace cult rooms were oriented towards the rising sun. Shaw (1977) was the first to measure the orientation of the palace. Blomberg and Henriksson (2001 and refs. therein) showed that in the equinoxes, because of the presence of the Aelias ridge on the eastern side of the Knossos palace, the rising position of the sun is shifted southward and was at (az 97.2 deg, alt 10.4 deg; for the upper limb of the sun). At the equinoxes, the rays of the rising sun enlightened a shallow stone bowl in the end the Corridor of the House Tablets, creating a reflection to the end wall, if the bowl were filled with water. Eleven days after the autumn equinox, the sun could be observed rising on the right corner of the door opening of the corridor. These kinds of observations could be used not only for observing the solar year, but also for keeping up a lunisolar calendar by adding an intercalated month every time a new crescent moon was observed in the evening during the 11-day period. Blomberg and Henriksson (2001) argued that in Knossos, the Minoan new year began in connection with the autumn equinox. Further connections to the sun's rising positions at Knossos at different times of the solar year were made by Goodison (2001; see Section 4.2.1 above).

However, the question remains, why was the orientation of the main axis of the palace not exactly to the equinoctial sunrises. The exact orientation of the palace axis in 2000 BCE was to the sunrise five days before the vernal equinox (i.e., the equinox day being the sixth), when the upper limb of the sun was at (az 100.4 deg, alt 10.4 deg; see Blomberg and Henriksson 2001). At this the position, the light of the rising sun enlightened a platform in the Inner Sanctuary of the Throne Room complex.

The orientation was not as exact five days after the autumn equinox, but the enlightening effect was valid. Goodison (2001) observed that some of the Messara tholos tombs were oriented towards the sunrises around the equinoxes.

This 'epagomenal' orientation may have been related to the date of a festival in the end of the Minoan year. The Egyptian year, which included 360 plus 5 epagomenal days, had probably influenced the Minoan festival calendar. Moreover, the very practice of letting the rising sun enlighten a cult statue is similar to that of Egyptian temples. However, the orientation of E-W axis of Knossos means that the Minoan solar year would have started in the spring, which contradicts the evidence presented above that the Minoan year would have begun in the autumn. Perhaps the solar and lunar calendar had different starting dates.

According to Goodison (2001), all four doorways of the Throne Room had their own significant orientations (2001; see also Section 4.2.1 above). She observed that looking outwards towards the central court, the orientation of the second door from left is oriented towards the sunrise at late August/early September. Also most of the Messara tholos tombs show this orientation, which she calls "the times of the dead". This terminology alone, however, suggests that the orientation was towards another celestial object. The heliacal rising of Spica occurred above the Aelias ridge in Knossos at (az 84.5 deg, alt 10.5 deg) nine days before the autumn equinox in 2000 BCE. As mythology connects the star to Persephone, a chthonic goddess, and to beliefs concerning the netherworld, it is understandable why one of the orientations at Knossos would have been to the heliacal rising of Spica, or to the sunrise on the date of the heliacal rise. That the exact date, and thus the orientations of the tholos tombs can vary by a few days, can be explained by visibility conditions, which have a considerable effect on the exact date of the observation of a heliacal rise event.

The Minoans at Knossos, observing regularly the movements of the sun near the equinoxes, must also have been aware of other prominent celestial phenomena occurring in this direction. Venus, which regularly appears in its varying positions near the rising sun, was likely noticed, as it was held in important position also in contemporary Egypt and Babylonia. For example, in 2000 BCE, one lunar month after the vernal equinox, Venus (about -4.4 mag, near its brightest phase) rose at (az 99.5 deg, alt 10.5 deg). Venus then was at its extreme southern rising position during that particular 584-day cycle (see Aveni 2001:81 for the possible rising positions of Venus) and seemed to rise near the same position for several days. The same arrangement occurred again after eight years, because of the Venusian octaeteris (see *Gournia* below).

*Mallia*
The central court of the Mallia palace is the second largest after Knossos. In the east and the south, mountains surround the palace. In the south, there is the rather small but prominent peak of Profitis Ilias, which perhaps was the place of a peak sanctuary associated with the palace (Fitton 2002:77). The palace may have been deliberately oriented towards the peak. In the north, the sea is close.

The long axis of the central court of the Mallia palace is oriented roughly N-S, as in all known palaces. As in Knossos, the cult rooms were located in the west wing, facing east, which suggests a deliberate orientation towards a celestial event in the eastern horizon (Marinatos 1934).

The E-W axes of the central court is directed towards the azimuth 108 deg, which is the location of a pass between two mountains in the east. After that point, the horizon level rises towards the south.

Blomberg and Henriksson (2001; 2005) measured the orientation of the axis of symmetry of the main cult room in the west wing of the palace at Mallia to be (az 107.6 deg), and argued that the sun rose between the mountains one lunar month after the autumn equinox in 2000 BCE, marking the time for the beginning of the Minoan agricultural year.

*Phaistos*

It is difficult to precisely measure the orientation of the central court of the palace of Phaistos, since the SE part of the court is completely gone, as well as half of the western part of the palace, as a result of an ancient earthquake. The E-W orientation measured for the central court of the New Palace is (az 93 deg).

In the north, the central court of Phaistos is directed towards the highest peaks of the mountain Ida, where also the Kamares cave is situated (Scully 1979). The palace seems to have had a special relationship to the cave instead of any peak sanctuary. In the east, there is the Messara plain and the horizon is very low.

Blomberg and Henriksson (2006) argued that the orientation of the central court of the palace had changed from the Old to the New Palace Period. The orientation of the old palace corresponded to the equinoctial sunrises around 2000 BCE.

The E-W orientation of the New Palace corresponds to the sunrises five days before the vernal equinox or five days after the autumn equinox in 1700 BCE, as the orientation of the Knossos central court and some of the Messara tholos tombs (see *Knossos* above).

*Zakros*

Zakros is the easternmost of the four largest palaces. Everywhere except in the east, the palace is surrounded by mountains. The orientation of the central court and the palace notably differs from the close N-S orientation of the central courts of Knossos, Mallia, Phaistos, and Gournia.

The cult rooms in the west wing of the palace of Zakros were oriented towards the southernmost moonrise (az 127 deg) (Shaw 1977; Henriksson and Blomberg 1996). There is archaeological evidence that a moon deity may have been important in the peak sanctuary of Petsofas (Blomberg and Henriksson 1996), which functioned under the administration of a palace on eastern Crete, possibly Zakros, Petras, or Palaikastro. However, the peak sanctuary closest to Zakros, Traostalos, seems to have been a place of solar worship (see Section 3.1).

*Gournia*

Mountains surround Gournia on three sides. In the north, there is the Mirabello bay. In the east, the great court is oriented towards a passage between mountains. In the south and in the west, the mountains rise high.

The orientation of the main court of Gournia is roughly along the N-S line. The court is connected to the "Governor's House", a small palace, which was built around 1600 BC (see, e.g., Fitton 2002:141-43). The town, however, functioned from EM to LM III.

Obtaining an exact measurement of orientation for the court is difficult, since it is not strictly rectangular. The town plan of Gournia was much affected by the hill it was placed on and is rather irregular. The great court was placed outside the palatial building.

The E-W orientation of the great court is between (az 85 deg), measured using the western side of it, to (az 79 deg), measured using the estimated eastern side. In 1600 BCE, Spica rose heliacally at (az 85 deg, alt 7 deg). The dates near this heliacal rise event correspond to the time of the solar orientation for "the times of the dead" as termed by L. Goodison (see *Knossos* above).

The orientation (az 79 deg) is towards the sunrise about one lunar month before the autumn equinox, or after the vernal equinox, in 1600 BC. Blomberg and Henriksson (2008) obtained this solar orientation for two Minoan shrines in Gournia, oriented to 78.9 deg ±1.0 deg and 78.3 deg ±0.6 deg, and built with a 300 years separation.

The position (az 80 deg, alt 7 deg) was also the rising point of Venus in its conjunctions with Spica a few days before the autumn equinox in 1700-1600 BCE. The conjunction Venus and Spica repeated itself every eight years with an error of just one day. Therefore, observing the heliacal risings of Spica, or just the sunrises near the autumn equinoxes, it would have been hard for the Minoans *not* to detect the eight-year cycle of Venus.

The results of Blomberg and Henriksson (1996; 2005 and refs. therein), as well as the interpretation of the calendrical disk in Section 3.2.2, indicate that the Minoans were aware of the lunar synodic octaeteris. The Venusian synodic octaeteris could have been recognized, too. However, since both the lunar and the Venusian octaeteris shifts, they should have been calibrated using the sun, or a notable star. Of all the stars brighter than 1.0 mag and observable in Crete, the most suitable star was Spica, since its heliacal rising date was nearest to the autumn equinox. Moreover, mythology closely links Spica to an important autumnal festival of two goddesses of Minoan origin (Section 2.4.3).

During its 584-day synodic period, Venus appears as a morning star for 263 days, disappears for 50 days, appears as an evening star for another 263 days, and disappears for 8 days (during which it is in front of the sun). This period was important for many ancient peoples, including the Babylonians and the Maya. Five of the synodic periods of Venus form an octaeteris, which differs only 2.3 days from eight tropical years and 3.9 days from 99 lunations. This is one directly observable Venusian octaeteris; the other is the eight year period of eight consecutive long and short empirical sidereal periods (ESI) of Venus (Aveni 2003 and refs. therein).

The observation of the octaeteris of the conjunction between Spica and Venus was possible because of the ESI of Venus. An ESI is the number of days elapsed between the consecutive passages of a planet in prograde motion by a given celestial longitude. It is believed that most ancient cultures were not aware of the sidereal periods of the planets, since they are difficult to observe. However, as Aveni (2003) pointed out, the Venusian sidereal period can be observed in connection with heliacal rises. In our case, the octaeteris consisting of ESIs is simply the period from one conjunction of Venus rising with Spica to another in a certain position as defined by an astronomical orientation of a Minoan building.

The question of possible deliberate orientations of Minoan palaces to Venus is a complicated one, and the issue must be considered very tentative at this point. The possibility of deliberate Venus orientations in 2000-1600 BCE should be determined by the dates of the reconstruction of the palaces, which are not yet exactly known. Moreover, since Venus is near the sun, orientations to it are easy to find, and are easily confused with possible solar orientations. However, at the same time, observations of Venus are almost unavoidable to anyone observing the sun, and Venus has been an important part of many mythologies. Usually, a Venus deity is in close relation to the solar deity, and could even be the same one. They may have been related also in Minoan Crete.

The date of the heliacal rise of Spica shifted gradually, which eventually prevented its conjunctions with Venus from happening before the autumn equinox. In 600 BCE, Spica rose heliacally at the autumn equinox at sea level on Crete.

In Gournia, there was also a third, Mycenaean shrine, which was oriented towards the last rays of the sun at the autumn equinox, so that they would reappear only at the day after the spring equinox (Blomberg and Henriksson 2008). Some of the Mycenaean graves on the Mainland were oriented towards the sunsets in autumn and winter (see Blomberg and Henriksson 2005). As the Minoans preferred eastern orientations for their graves, the orientation of the shrine may be related to the Mycenaean beliefs concerning the afterlife.

*Petras*

The palace of Petras is the smallest of the six palaces considered in this paper. It is located on a north-protruding hilltop near the coast of Sitia bay, with a view down to the bay and the valley, which was probably governed by it. The valley is on the western side of the hilltop. In the south, the mountains rise high.

The central court, around which the palace was reconstructed after the destruction in the end of Middle Minoan II (1700 BCE), was once again rebuilt and reduced in dimension, although retaining the original orientation, in Late Minoan IB (1500 BCE).

The E-W orientation of the main axis of the Petras palace (az 57 deg) differs from the rather strict N-S/E-W trend presented by Knossos, Phaistos, Mallia, and Gournia. The central court of Petras was oriented to the northernmost moonrise (az 57 deg, alt 3 deg). This can be compared to Zakros, which was oriented towards the southernmost moonrise. Thus, the palaces of the easternmost Crete seem to have been oriented towards the moon rather than towards the sun.

## 4.3 Minoan 'sacred landscape'

As shown above, there is now accumulating evidence that the palaces and graves were deliberately oriented towards celestial events and sacred mountains, many of them having an important peak sanctuary on top of them, or, in the case of Phaistos, a sacred cave.

The bonfires that were burned in the peak sanctuaries were probably related to solar worship. The caves, in turn, were the places of worship of chthonic deities, one of which was the deity associated with the double axe. This division between the underworld and the sky, which was the dwelling place of celestial divinities, was reproduced in the architecture of the palaces, where the alteration of dark and light spaces was used in the rituals.

It is probable that the Minoans, as many other ancient peoples, perceived the world as having essentially three layers: the celestial realm, the underworld, and the human world in between.[7] The divinities that lived in the other two realms could be reached and pleased only by closely observing the celestial and other natural phenomena, and by connecting the human behaviour, i.e., the religious rites, with the symbolism and occurrence of those.

The Minoan religion as a whole had a strong aspect of fertility and periodical renewal. This close connection of human beings to the processes of nature manifested itself also in the identities of deities. The realm of the sea, governed by the "Earth-Shaker", had a strong life-giving side, as well as a life-destroying one. The importance of the sea is clearly seen in the Minoan art, especially in the Late Minoan. In some representations, a Minoan goddess is depicted rising from the ground - a scene, which has a direct relation to growing vegetation (see, e.g., Evans 1925, Fig. 16). Celestial symbols, too, are frequent in Minoan art dating from all periods. As the Palaikastro mould shows, the Poppy Goddess was not only a chthonic fertility goddess, but also the goddess of celestial cycles.

The Minoan, as, e.g., the Mayan art, is very rich in symbolism. The divinities are depicted in natural landscape as mistresses and masters of the animal and floral kingdoms, and all human constructions intermingle with the surrounding landscape both in shape and in color. Most of the symbols depicted in Minoan art, as in other contemporary art, are not merely decorations, but have a religious meaning – a fact that our own, secular time often seems to overlook. By the use of the sacred symbols, the architecture and art in the palaces were targeted to create a spiritual experience to a visitor by constantly reminding her or him of the presence of the divinity. Most of these symbols were natural phenomena, or objects that a Minoan person would meet in his or her everyday life. In effect, the Minoans lived their whole lives in a realm religious to the core.

It can be said that the Minoans, like the Maya, lived in a 'sacred landscape' around them. Many of their architectural choices were determined by the interplay between the human constructions and the natural landscape, including the celestial events. The palaces were 'embedded' in the sacred landscape, and the rituals performed in them replicated and, thus, connected with the events occurring in the upper and lower realms.

---

[7] Castleden (1990:133) saw the tripartite shrines as a manifestation of this division.

**4.4 Minoan festival calendar and the origin of the Eleusinian Mysteries**

Much of the Minoan religion seems to have been based on the ideas of renewal and resurrection. The legends of Demeter and Iasion, the periodically dying Cretan Zeus and Persephone the Virgo reflect the beliefs in the origin of the Minoan religious rituals. These beliefs had their obvious origins in the cycles of the natural world: the animal life cycles, the annual vegetation cycle, and the periodical movements of the celestial bodies.

The Greek mythology and the archaeological and archaeoastronomical evidence together suggest that the most important rituals of the Minoans were performed on calendrically important dates in the palaces. The rituals included sacrifices, communal feasts, ecstatic dancing, and possibly a *hieros gamos*. They also included the use of light and darkness in the epiphanies of the primary goddess, the Poppy Goddess. The legend of Minos *enneoros* suggests that, at least in the Mycenaean era, the need to the ritual renewal of the power of the regent was included in the religion. This ritual was required every eight years, probably in connection with the end of the lunar octaeteris.

The orientations of the Minoan buildings were affected by the dates of their ritual calendar. The palaces of Knossos, Phaistos, Mallia and Gournia were oriented to the rising sun, whereas Zakros and Petras were oriented to the rising moon. These orientations were related to the lunisolar calendar and, thus, to the solar and lunar deities.

The orientations of the palaces to lunisolar calendrical events were governed by the Poppy Goddess, whom the Palaikastro mould shows to have been a solar and calendrical deity (see Section 3.1). Thus, the solar deity of the Minoans was female. This great sun goddess of the Minoans was the original Demeter, a chthonic, fertility and calendrical goddess, who had poppies as her symbols. It is even possible that her name already was Demeter, *DA-MA-TE* in Linear A (Owens 1996a). Thus, she can be called the 'Minoan Demeter'.

The orientations of the palaces of Knossos and Phaistos were towards the sunrise five days after the autumn equinox, or before the vernal equinox. I suggest that this was the time of an important festival related to the Minoan New Year and, perhaps, to the end of the lunisolar calendrical cycle every eight years. On these dates, the rising sun enlightened the platform in the Throne Room complex, where likely the statue of a deity was kept. This resembles the practice in the temples of Egypt. The model for the annual 360-day calendar and the New Year's festival probably was an Egyptian one.

It is difficult to say, whether the New Year's festival was celebrated in the spring or in the autumn. The orientation of the palace of Knossos points to the 360-day period having ended five days before the vernal equinox. Marinatos (1986:58-9) suggested that there was an important spring festival related to rituals of renewal, and to the number eight, as revealed by the eight paintings in room 4 of the West House at Akrotiri, Thera. However, Blomberg and Henriksson (2001) argued that the orientation of the Corridor of the House Tablets in Knossos could be used to regulate the lunisolar calendar in connection with the autumn equinox only.

In addition to the New Year's orientation, there were several other orientations to celestial events at Knossos. The midsummer, midwinter and equinoctial sunrises, as well as their relation to the lunisolar calendar were all recorded in stone. The Knossian palace was, in effect, like a giant calendar, which preserved and presented the basic Minoan calendrical cycles.

One of the orientations of the Throne Room was to the heliacal rising of Spica. This orientation

to "the times of the dead", seen also in the orientations of the Messara tholos tombs (Goodison 2001)and the central court of Gournia, was likely related to the date of one of the most important religious festivals of the Minoans.

As seen in Section 2.4.3, mythology connects Spica to the resurrecting vegetation goddess Persephone, and its heliacal rising date at the time of the grape harvest to Dionysos. Because both of these deities had a strong chthonic aspect, the Minoan festival celebrated at this time of year was likely related to the cult of the dead. Both of these deities were also part of the Eleusinian mysteries. The time of Eleusinian mysteries would have coincided with the heliacal rising of Spica in early September in Minoan times. Therefore, this particular Minoan festival was likely a predecessor of the Eleusinian Mysteries.

Besides the Eleusinian Mysteries, the other great festival related to Demeter in Classical times was Thesmophoria. Thesmophoria was celebrated about one lunar month after the Eleusinian mysteries, which corresponds to the present September-October. The mention of Hesperos, Venus as the evening star, appearing before Thesmophoria, shows that the start of the festival was originally connected to planet Venus in some way. This may reflect the Minoan observations of Venus rising close to the sun or Spica near an autumnal festival of the great goddess.

Thesmophoria was associated with Demeter alone, and the Eleusinian mysteries with both Demeter and Persephone. Both festivals had a close connection to the cycle of vegetation. For Minoans, Demeter as the great goddess was probably associated with the equinoxes and the start of the octaeteris. Persephone as a chthonic deity is more closely connected to the invisibility periods and subsequent 'resurrections' of Venus, Spica and the moon. However, the question of the origin of Persephone, the daughter of Demeter, is a difficult one, since it is impossible to say for certain, whether there was a Minoan daughter goddess. It may just as well be that the Minoan solar goddess had two aspects, the chthonic and the celestial one, and that these were worshiped separately. The Hittites had separate sun and Venus goddesses, but two aspects of the same great goddess: the sun goddess Arinna and Hepat (see, e.g., Akurgal 1962:75-81).

In one Minoan-Mycenaean scene, the Poppy Goddess is depicted rising from the ground towards a male attendant, just like Persephone in Classical times (Evans 1925, Figs. 16-18). On the Hagia Triada sarcophagus from LM IIIA, two goddesses are riding on a chariot (Nilsson 1950:427, 440). The scene much resembles representations of the chariot of the sun common in the Bronze Age. Perhaps they are the solar goddess and the Venus goddess, who move close to each other in the sky. In the Phrygian cult, the goddess rode on a chariot (Ramsay 1912:71), which indicates that one of the goddesses on the Hagia Triada sarcophagus was the same as the Phrygian one, i.e., Demeter.

The significance of the original, Minoan Demeter was probably first altered when she was transferred into the Mycenaean system of beliefs. A more profound change must have happened, when she became part of the Dorian Greek pantheon, which had a dominant male god, as well as a male solar deity. If the original Minoan system was close to the Anatolian one, then the Minoan pantheon merging with the Greek one resulted in Venus replacing the role of the sun as Demeter's 'star'. The process may have resulted also in the splitting of the identity of the Minoan Demeter, when she became fused with the less powerful corresponding Greek goddesses. In this way, the Minoan observations of Venus appearing with the sun every eight years probably eventually contributed to the mythology of Demeter and Persephone. Similarly, the heliacal rising of Spica near the autumn equinox and the conjunctions of Spica with Venus could be Demeter meeting Persephone after her annual visit to the netherworld. The exact relations of the myths to the celestial events cannot be reconstructed, but some kind of derivation of the Classical relationships between these goddesses from the movements of the celestial bodies, based on their significance in the Minoan observing tradition, is probable.

As discussed in Section 2.4.2, one original companion of the great Minoan goddess was a young male, the Cretan Zeus or Dionysos, who perhaps was associated with the moon. The lunar and Venusian octaeterides, followed using the autumn equinox and the conjunctions with Spica, could thus have had their mythological representations in the form of the male lunar deity and the periodically dying goddess, respectively. However, the goddess figures found in Petsophas suggest a female lunar deity (Blomberg and Henriksson 1996). Perhaps the Minoan male fertility deity was associated with the weather, like the Hittite Teshup, rather than celestial bodies.

The ceremonies carried out in the Throne Room of Knossos at the autumn festivals likely are the origin of the mystery rites performed in Eleusis and in other later mystery cults. In the Phrygian, Eleusinian-like mystery cult of Demeter and the moon god Mên Askaënos, the initiation rites took place in a temple very much like the Knossian Throne Room, with a "throne" seat and a "lustral basin" (Hardie 1912; Ramsay 1912; Evans 1921:4-5). A *hieros gamos* between the goddess and the god, represented by the priestess and the priest, was part of the ceremony (Ramsay 1912:51-4). The cult of Phrygian Attis had a torch-bearer, which means that it also had a procession like the Eleusinian Mysteries (Ramsay 1912:73). The way the Phrygian torch-bearer carried his dress on his shoulder brings to mind the leader of the procession depicted on the Minoan "Harvester vase", who carries a garment decorated with apparent crescent symbols on his shoulder (see, e.g., Fitton 2002, Fig. 87; see also: Marinatos 1986:51-72 on the symbolism of Minoan sacred garments). Moreover, the symbolisms of the crescent, the rosette, the ear of corn and the number eight were also present in the Anatolian cult (Hardie 1912:134; Ramsay 1912:75-6). It is therefore likely that in the origin of both the Eleusinian and the Phrygian cult there was an earlier Minoan cult of Demeter and her companion, and that both had preserved some key aspects of the Minoan rituals and symbols.

The legend of Minos renewing his divine laws on every ninth year of his reign may have its direct origin in the Minoan practice of a periodically reigning priest-king. This original Minos then would have been the person to play the part of the lunar god in the sacred ceremonies of the sun goddess. In Mycenaean Knossos, Minos would have been the most important person, the earthly ruler, *wanax*. But the question remains, how much did the Minoan governmental system, religion and rituals differ from the later Minoan-Mycenaean one.

There is an arc resembling the moon sickle depicted on the Knossian throne seat (see Fig. 895 in Evans 1935). The griffin on the west wall, however, points to a female having been seated on the throne, as in Mycenaean Pylos (Castleden 2005:167, 193). The "Throne Room" only occurs at Knossos and may have been a late addition (Hitchcock 2003), which may indicate the introduction of the Mycenaean sacral kingship in the Late Minoan. Whatever the case, the religious imagery shows that the primary deity was female in Mycenaean Knossos, too.

Although the goddess was more important than the god, the high priestess did not hold the highest political power in Mycenaean times. There is evidence of an important female, the "key-bearer", who was a priestess, but it is clear she did not have power over other political entities (Castleden 2005:81). At present, it cannot be concluded whether the Minoan high priestess had had any significant political relevance before the arrival of the Mycenaeans. However, the conflicts concerning claims in land-ownership, which apparently sometimes arose between a priestess living in the palace of Pylos and the officials in the city (Hooker 1995:19), may hint at the former Minoan practice where the priestess had held the highest political power. The Mycenaean rule would then not have greatly altered the structures of the former Minoan governmental system, but merely transferred the power to the Mycenaean elite.

In conclusion, it is suggested that before the Minoan-Mycenaean period and possibly also during it, the primary deity of the Minoans was a solar goddess, the original Demeter. She had two aspects or an important female companion, and a subordinate male, possibly a lunar god, as her consort. She had her "mysteries", i.e., communal rituals related to death and regeneration. The palaces were her temples, used in her epiphanies, which were performed by priestesses. The dates of the religious festivals were determined from the movements of the celestial bodies, and the ceremonies carried out in them had much in common with the later Eleusinian and Phrygian Mysteries, likely including a *hieros gamos*. In the festival rituals, the high priestess and the priest would act in the epiphany of the great goddess and her male companion, replicating the cycle of nature and the celestial bodies in the act of legitimization of their rule. The orientations of the Minoan temple-palaces and graves towards the east can be seen as a manifestation of the same Minoan beliefs concerning the afterlife: the renewal of the natural world was replicated both in the myth of a fertility goddess and her periodically resurrecting companion, and the fate of a human being. The power of the Minoan religious administration was based on ruling this connection between the layman and the gods. The rulers of the palaces possessed a special connection with the divine and sacred knowledge on rituals, including the ability to follow the calendrical cycles.

## 5. Summary

Literary, archaeological and archaeoastronomical evidence on Minoan astronomy and calendrical system has been examined. It is concluded that the Minoan society had religious beliefs and ritual practices, which were closely related to periodical celestial events.

In Minoan art, symbols for celestial objects are frequently depicted, often in clearly religious contexts. The most common are various crosses, spirals and rosettes, which are identified as solar and stellar symbols. The palace of Knossos was amply decorated with these symbols.

A clay disk with an astronomical motif depicted on it has been identified as a Minoan-Mycenaean ritual object showing the most important calendrical cycles, especially the 99-month lunar octaeteris. The disk has been compared with the large Minoan stone *kernoi*, and it is concluded that both types of disk likely had a ritual purpose related to the calendrical system used by the Minoans. These kinds of ritual calendrical objects were most likely used in the palaces and shrines.

Orientations of Minoan buildings and graves have been examined. The central courts of Knossos, Phaistos, Mallia, and Gournia were oriented to the rising sun, whereas the Eastern palaces Zakros and Petras were oriented to the southernmost and the northernmost risings of the moon, respectively.

The central courts of Knossos and Phaistos were oriented to the sunrise of five days before the vernal equinox or after the autumn equinox. On this date, the rays of the rising sun enlightened a platform, where a statue was likely kept, in the Inner Sanctuary of the Throne Room complex. The date was related to a Minoan festival celebrated on the five epagomenal days in the end of the solar year. The model for this kind of solar calendar probably came from Egypt.

In the Minoan religion, the idea of renewal was central. The myths of Minos and the two goddesses of Minoan origin, Demeter and Persephone include the cycles of eight, which can be related to lunar and Venusian periods. The tradition of Minoan astronomical observations of the sun, the moon, Spica, and to Venus and Spica in conjunction is reflected in these myths.

In Minoan Crete, Spica rose heliacally in early September, before the autumn equinox, marking the grape harvest. One of the orientations of the Knossian Throne Room, the western side of the great court of Gournia, as well as the orientations of the majority of Mesara tholos tombs were towards the heliacal rise of Spica or to sunrise near the date of the heliacal rise. This was probably the time of an important Minoan religious festival, a predecessor of the Classical Eleusinian mysteries.

The religious rituals celebrated in the palaces made use of the alteration of light and darkness, and re-examination of the archaeological evidence available suggests that also reflective properties of materials were made use of. The ceremonies utilizing light and reflection were probably connected to the epiphanies of deities, most importantly the Poppy Goddess, who was a female solar deity, the 'Minoan Demeter'.

| Name | Short ax. az (deg) | Long ax. az (deg) |
|---|---|---|
| Knossos | 100 | 190 |
| Phaistos | 93 | 183 |
| Mallia | 108 | 198 |
| Gournia | 79-85 | 169-175 |
| Petras | 57 | 147 |
| Zakros | 127 | 217 |

**Table 1**. Orientations of the axes of the central courts of six Minoan palaces.

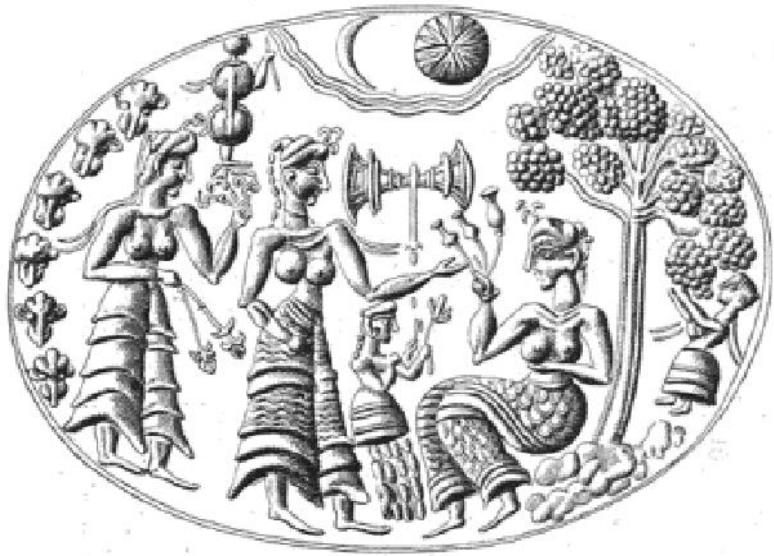

**Figure 1.** Minoan "Poppy Goddess" seated under a tree and greeted by other goddesses. Scene on a Minoan-style gold ring from Mycenae. After: A. Evans 1921, Fig. 4.

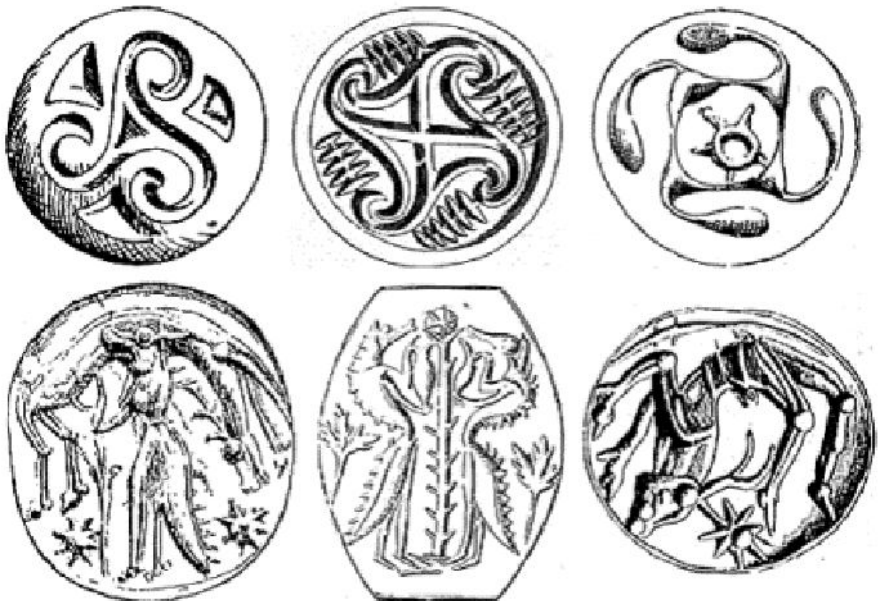

**Figure 2.** Spirals and solar/stellar motifs on Minoan seals. After: A. Evans, *The Palace of Minos,* Vol. II:1, Fig. 106a; Vol. IV:1, Fig. 255c; Vol. IV:2, Figs. 419, 364, 377, 449.

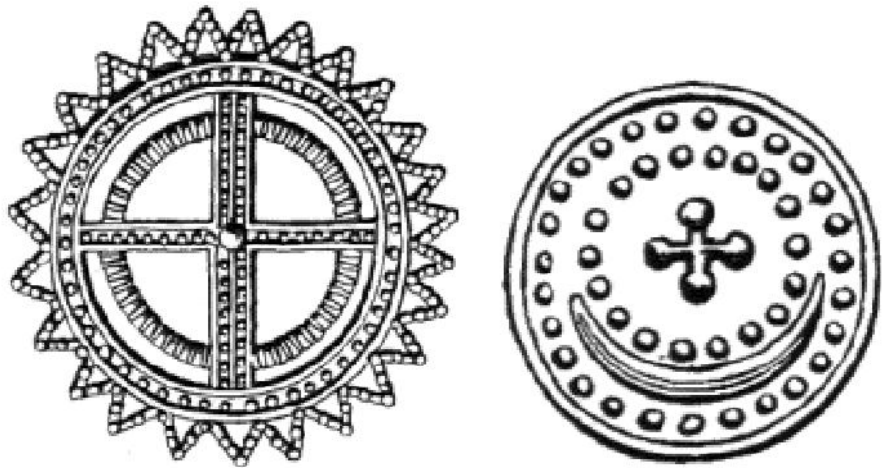

**Figure 3.** 'Solar wheel' and an object resembling the Minoan cup-holed *kernoi*, originally depicted with the Poppy Goddess on a mould from Palaikastro. After: A. Evans, *The Palace of Minos,* Vol. I, p. 514, Fig. 371.

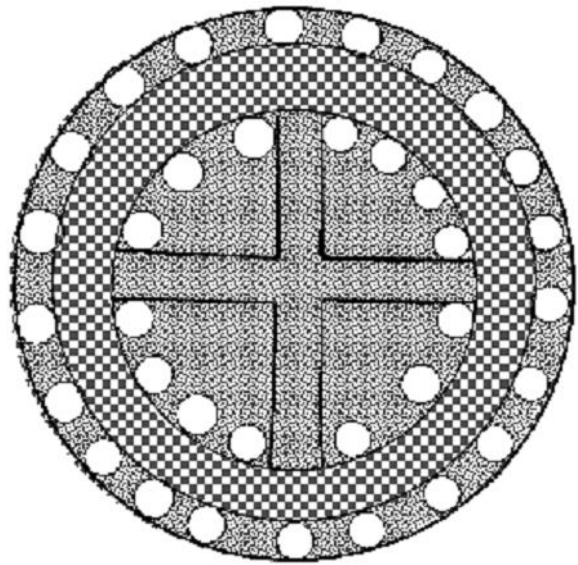

**Figure 4.** Object depicted on a MM IA jug from Knossos. Redrawn by M.R., after A. Evans, *The Palace of Minos,* Vol. IV:1, Fig. 61.

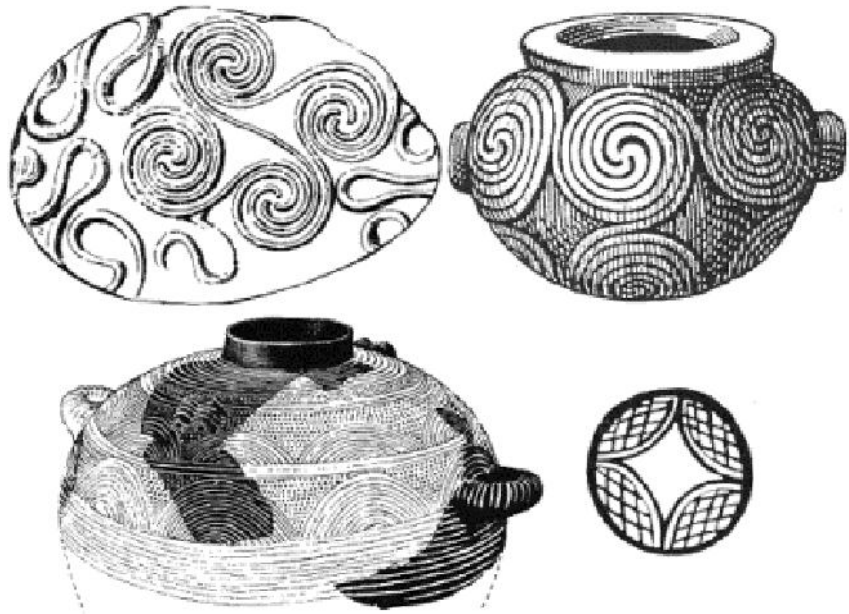

**Figure 5.** Spirals and *vesicae piscis* in Early Minoan pottery. From upper left to lower right: bottom of a statue of a dove with two youngsters from a Messara tholos tomb (EM III); stone bowl from Platanos (EM III); jar from Vasiliki (EM I); *vesicae piscis* –type stellar motif on a pot sherd from Gournia (EM III). After: A. Evans, *The Palace of Minos,* Vol. I, Fig. 86; Vol. II:2, Fig. 104a; Vol. I, Figs. 22, 80b.

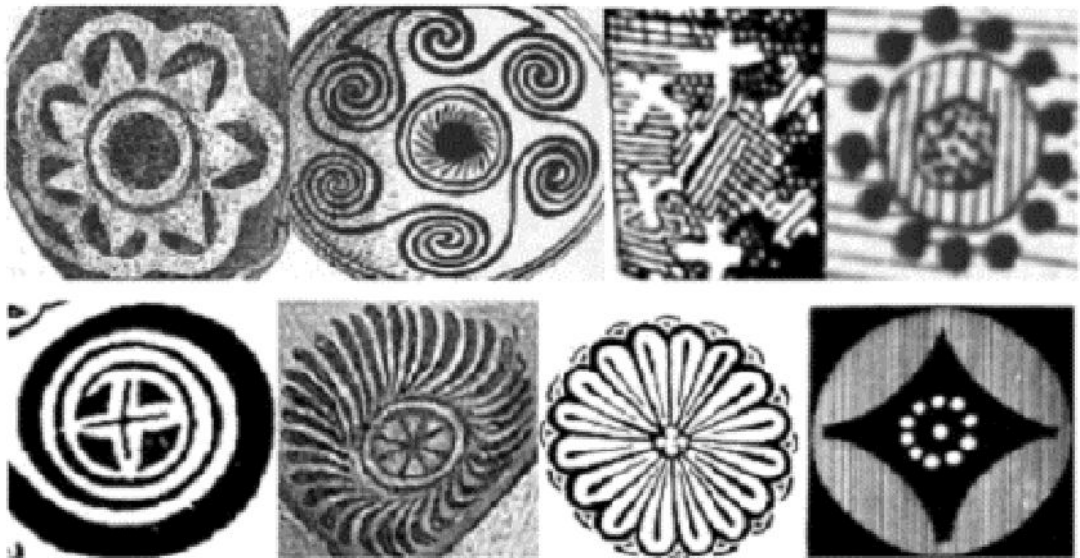

**Figure 6.** Solar and stellar motifs in Minoan pottery. From upper left to lower right: "pilgrim's flask" from Zakros (MM IIIA); another from Palaikastro (LM IA); tumbler vase from Knossos (MM IA); bowl from the town of Knossos (MM IA); *pithos* from Knossos (LM II); pedestal bowl from the Knossian Temple Tomb (LM III); pithos from Knossos (LM II); a Minoan ceramic pattern (MM IB). After: A. Evans, *The Palace of Minos,* Vol. II:1, Figs. 121b-c; Vol. IV:1, Fig. 64; Vol. II:1, Fig. 205; Vol. IV:1, Fig. 271; Vol. IV:1, Fig. 306; Vol. IV:1, Fig. 286; Vol. I, Fig. 194h.

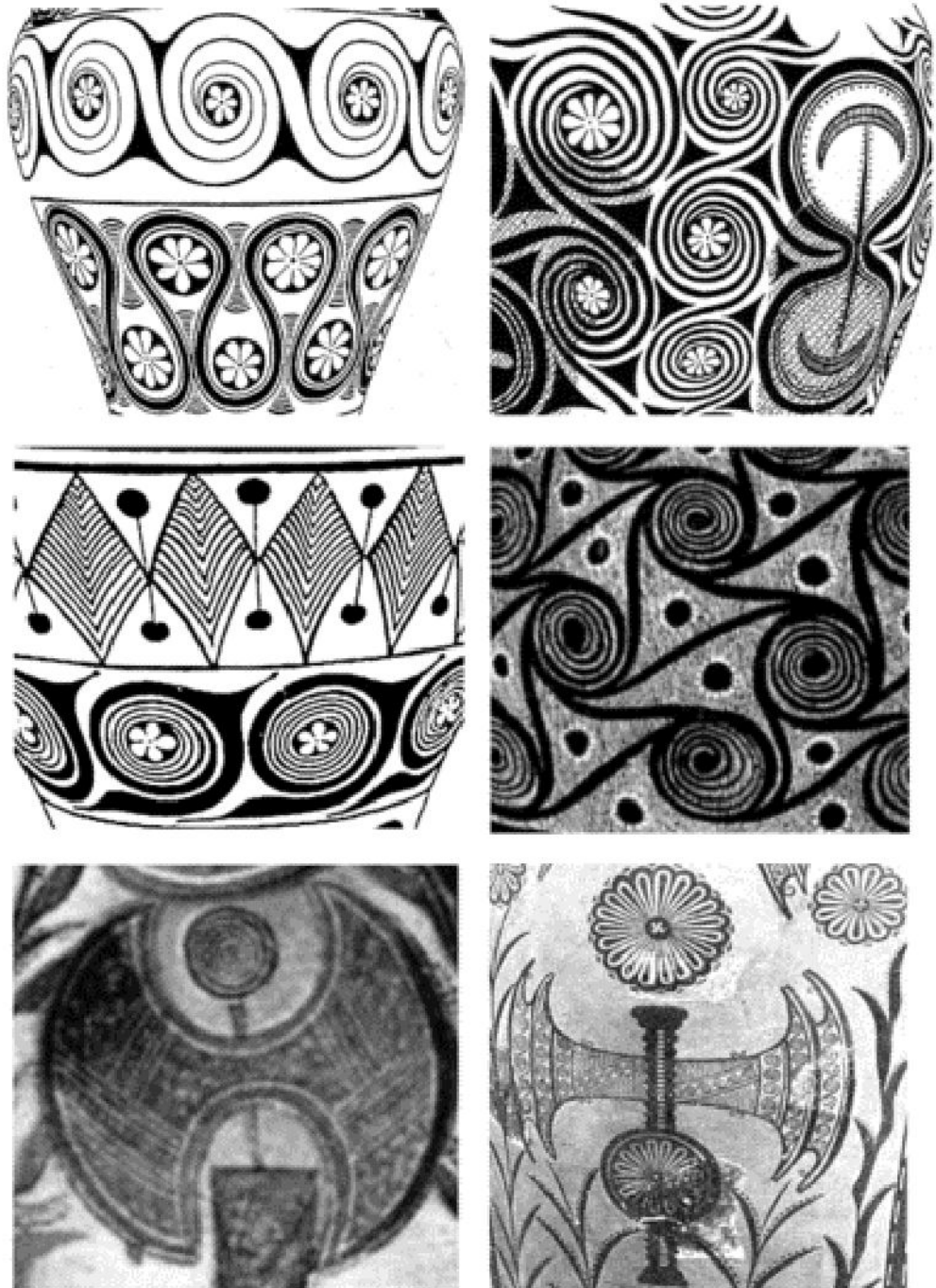

**Figure 7.** Spiral, rosette and double axe motifs on *pithoi*. From upper left to middle left: three Knossian "palace style" *pithoi* (LM II). Middle right: *pithos* from Knossos (LM IA). Lower left: *pithos* from Pseira (LM IA). Lower right: "palace style" pithos (LM II) from Knossos. After: A. Evans, *The Palace of Minos,* Vol. IV:1, Fig. 282; Vol. III, Fig. 199; Vol. IV:1, Fig. 260; Vol. II:2, Figs. 245, 284; Vol. IV:1, Fig. 285.

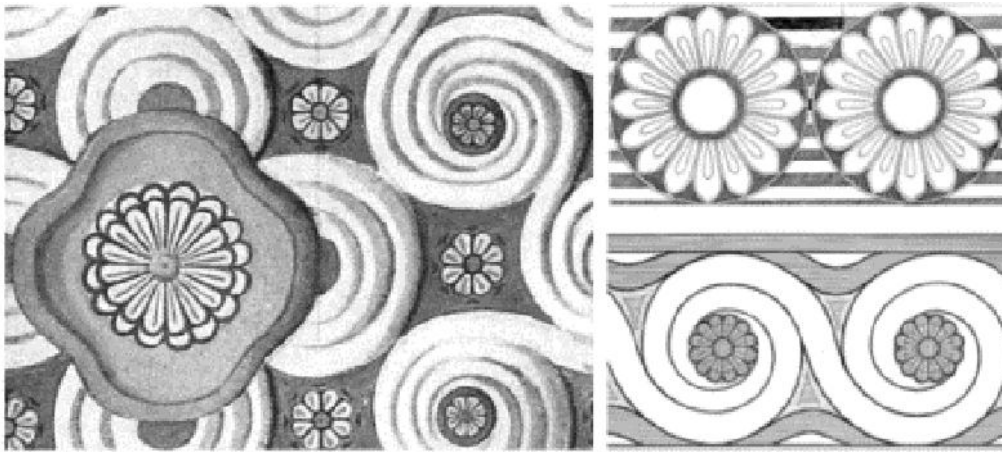

**Figure 8.** Spiral and rosette motifs on Knossian decorative patterns. From left to lower right: painted stucco ceiling from Knossos; restoration of crystal inlays found in the Western Temple Repository of Knossos; dado from Hall of the Double Axes. After: A. Evans, *The Palace of Minos,* Vol. III, Plate XV; Vol. I, Fig. 344a; Vol. III, Fig. 299.

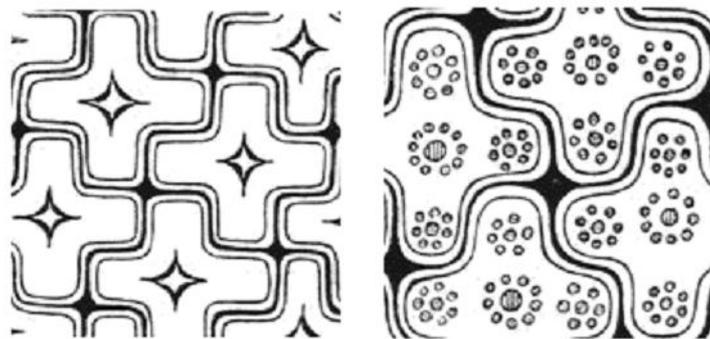

**Figure 9.** Patterns on Minoan garments from the processional fresco of Knossos (MM III). After: A. Evans, *The Palace of Minos,* Vol. II:2, Fig. 456d-e.

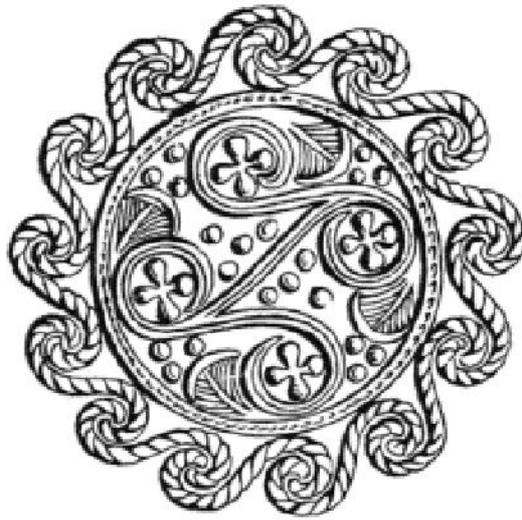

**Figure 10.** Knossian seal impression from MM I period. After: A. Evans, *The Palace of Minos,* Vol. II:1, Fig. 111.

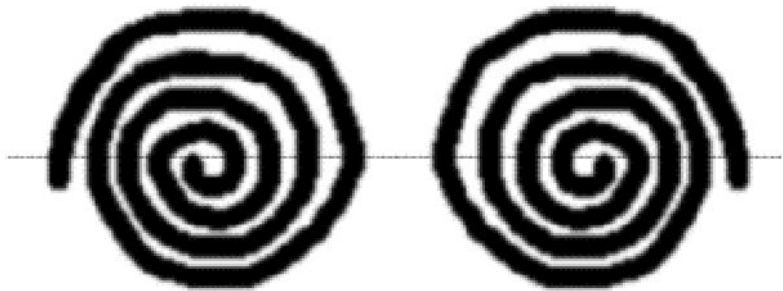

**Figure 11.** The simple spiral form represents the movement of the sun on the sky from summer solstice to winter solstice (the spiral on the left) and vice versa (the spiral on the right). The ancients believed that the sun followed a path similar to its daily path during the night in the underworld (the lower halves of the spirals in the figure). The spirals have been drawn using the rock carvings of the Knowth megalithic mound as models.

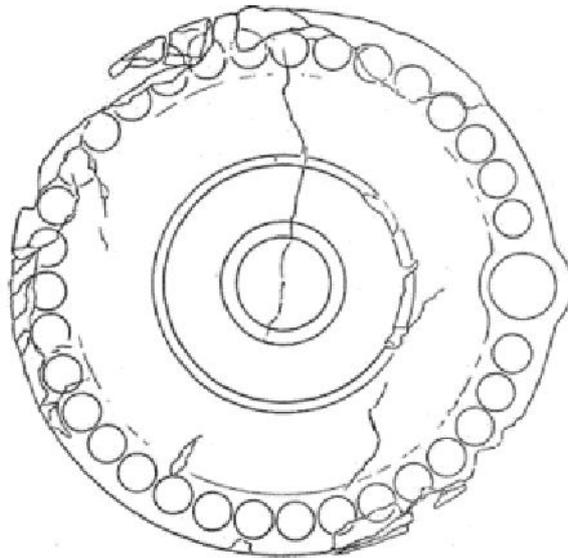

**Figure 12.** Stone *kernos* of Mallia (see Section 3.2.1). After: A. Evans, *The Palace of Minos,* Vol. III, Figure 263.

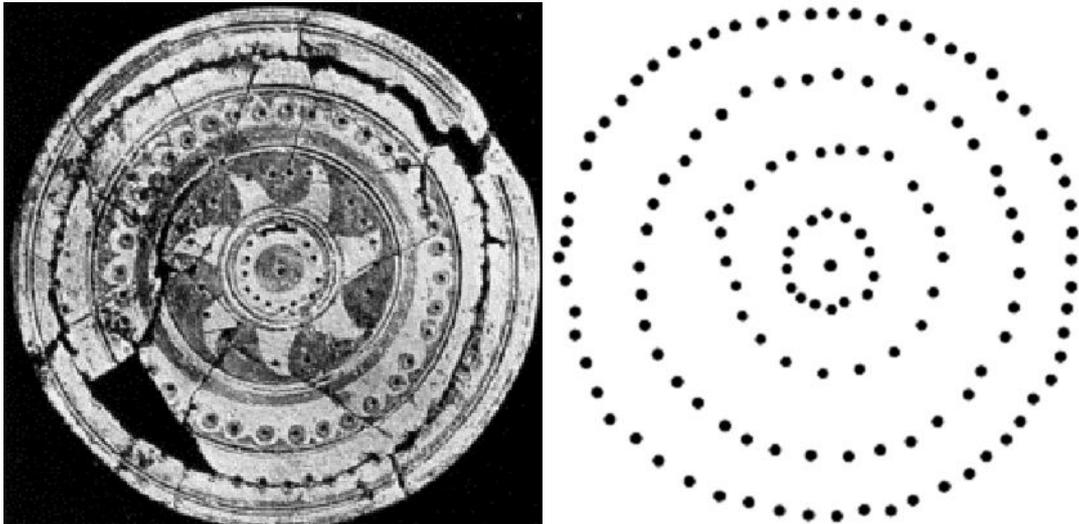

**Figure 13.** Left: perforated clay disk from LM III, after: A. Evans, *The Palace of Minos,* Vol. IV:2, Figure 720c (see Section 3.2.2). Right: the four concentric circles of holes in the disk.

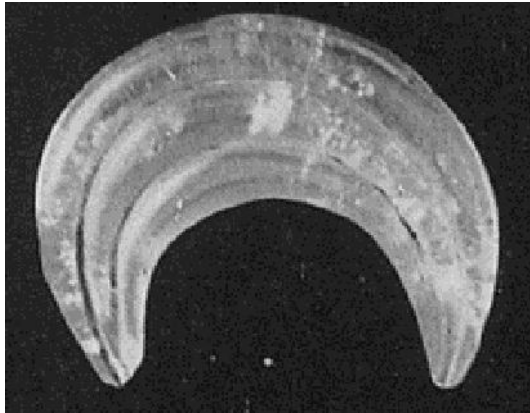

**Figure 14.** Crescent-shaped crystal found on the floor of the Knossian Throne Room. After: A. Evans, *The Palace of Minos,* Vol. IV:2, Figure 900.